\documentclass{aa}  

\usepackage{graphicx}
\usepackage{xcolor}
\usepackage{multirow}
\usepackage{lscape}
\usepackage{threeparttable}

\usepackage{txfonts,textcomp}

\usepackage{natbib,twoopt}
\usepackage{diagbox}
\usepackage[colorlinks=true, linkcolor=blue, citecolor=blue, urlcolor=blue, breaklinks=true]{hyperref} %% to avoid \citeads line fills
\bibpunct{(}{)}{;}{a}{}{,} %% natbib format for A&A and ApJ
\makeatletter
\newcommandtwoopt{\citeads}[3][][]{\href{http://adsabs.harvard.edu/abs/#3}%
{\def\hyper@linkstart##1##2{}%
\let\hyper@linkend\@empty\citealp[#1][#2]{#3}}}
\newcommandtwoopt{\citepads}[3][][]{\href{http://adsabs.harvard.edu/abs/#3}%
{\def\hyper@linkstart##1##2{}%
\let\hyper@linkend\@empty\citep[#1][#2]{#3}}}
\newcommandtwoopt{\citetads}[3][][]{\href{http://adsabs.harvard.edu/abs/#3}%
{\def\hyper@linkstart##1##2{}%
\let\hyper@linkend\@empty\citet[#1][#2]{#3}}}
\newcommandtwoopt{\citeyearads}[3][][]%
{\href{http://adsabs.harvard.edu/abs/#3}
{\def\hyper@linkstart##1##2{}%
\let\hyper@linkend\@empty\citeyear[#1][#2]{#3}}}
\renewcommand*\aa@pageof{, page \thepage{} of \pageref*{LastPage}}

\begin{document} 

   \title{Broad line region echo from highly accreting intermediate-mass black hole candidate SDSS J144850.08+160803.1}
   \subtitle{First probe of intra-night variability \&\ reverberation mapping}

   \author{Mariia Demianenko\inst{1\thanks{Fellow of the International Max Planck Research School for Astronomy and Cosmic Physics at the University of Heidelberg (IMPRS-HD).},2,3},
          Anton Afanasiev\inst{4,5},
          Evgenii Rubtsov\inst{5},
          Victoria Toptun\inst{6},
          Jörg-Uwe Pott\inst{1},
          Alexandr Belinski\inst{5},
          Franz E. Bauer\inst{7},
          Igor V. Chilingarian\inst{3,5},
          Kirill A. Grishin\inst{8,5},
          Marina Burlak\inst{5},
          Natalia Ikonnikova\inst{5}
}

   \institute{Max-Planck-Institut für Astronomie, Königstuhl 17, 69117 Heidelberg, Germany,
              \email{demianenko@mpia.de}
         \and
     Department for Physics and Astronomy, Heidelberg University, Im Neuenheimer Feld 226, 69120 Heidelberg, Germany
         \and
    Center for Astrophysics --- Harvard and Smithsonian, 60 Garden Street MS09, Cambridge, MA 02138, USA
        \and
    LESIA, Observatoire de Paris, Université PSL, CNRS, Sorbonne Université, Université Paris Cité, 5 place Jules Janssen, 92195 Meudon, France
        \and
    Sternberg Astronomical Institute, Moscow M.V. Lomonosov State University, Universitetskij pr., 13,  Moscow, 119234, Russia
        \and
    European Southern Observatory, Karl-Schwarzschildstr. 2, 85748, Garching bei M\"unchen, Germany
        \and
    Instituto de Alta Investigaci{\'{o}}n, Universidad de Tarapac{\'{a}}, Casilla 7D, Arica, Chile
        \and
    Universit\'e Paris Cit\'e, CNRS(/IN2P3), Astroparticule et Cosmologie, F-75013 Paris, France
\\
             }

   \date{Received 2025; accepted 07/08/2025}

\abstract
    {Elusive intermediate-mass black holes (IMBHs; $100~M_{\odot} \leq M_{\rm BH}\leq 2\times10^{5}~M_{\odot}$) can be used as ``time-squeezing'' machines, enabling studies of active galactic nucleus (AGN) geometry via reverberation mapping (RM) on much shorter timescales than their supermassive siblings.}
  % aims heading (mandatory)
    {Constraints on the broad line region (BLR) radius for IMBH candidates across a broad range of Eddington ratios help probe the unexplored faint end of the radius-luminosity ($R-L$) relation in AGNs. This development opens up the opportunity to build a more robust $M_{\rm BH}$ estimator. The present study is aimed at: (a) confirming a highly accreting IMBH candidate and (b) demonstrating the feasibility of the first photometric BLR RM campaign for IMBHs with high Eddington rates.}
    {SDSS J144850.08+160803.1 (J1448+16) was identified as an IMBH candidate from a broad H$\alpha$-selected spectroscopic sample from SDSS. We carried out \textit{XMM-Newton} X-ray observations to confirm its AGN status, along with narrowband H$\alpha$ and broadband SDSS~g\'~monitoring over five months (March--July 2024) using a 60-cm telescope at the Caucasus Mountain Observatory. These time series allowed us both to probe the short-timescale variability and extract the time lag between the BLR and accretion disk (AD) continuum.} 
    {\textit{XMM-Newton} detected J1448+16 as a bright X-ray point source with a photon index of $\Gamma = 2.32^{+0.15}_{-0.13}$ and X-ray luminosity of $L_{2-10\,\rm{keV}} = (3.3^{+0.5}_{-0.4}) \times 10^{41}$\,erg\,s$^{-1}$, confirming its AGN activity. From the SDSS optical spectrum and X-ray properties, we estimated a  BH mass of $\sim (0.9-2.4)~\times10^{5} M_{\odot}$ and Eddington rate of $\sim 37-112\%$. 
    We report high-amplitude $\sim 55\%$ intra-night ($\sim 1.7$~h) H$\alpha$ variability for this highly accreting IMBH and extract a tentative BLR RM radius estimate of $\sim 1-8~\mathrm{days}$.} 
    {This work offers a proof of concept for further high-Eddington-rate IMBH variability studies and BLR RM campaigns, which will be essential for an efficient calibration of the $R-L$ relation at the faint end.}

   \keywords{Galaxies: active -- Galaxies: dwarf -- Galaxies: nuclei -- Galaxies: Seyfert -- Accretion, accretion disks}
   \titlerunning{The first probe of reverberation mapping on highly accreting IMBH}
   \authorrunning{Demianenko et al.}
   \maketitle
%
%-------------------------------------------------------------------

\section{Introduction}

Active galactic nuclei (AGNs) are powered by massive black holes (MBHs) surrounded by accretion disks (ADs) and broad-line regions (BLRs) comprised of hot, dense, rapidly moving gas clouds \citep{Antonucci1993,Urry1995,elvis00}.
At larger radii from the center, clumpy tori are typically observed, with significant concentrations of small, micron- and submicron-sized dust grains \citep{2015arXiv150102001A}. Finally, colder and less gravitationally influenced gas clouds are located in larger kpc-scale narrow-line regions (NLRs), whose emission is characterized by strong forbidden line emission. Despite this basic structural uniformity, the wide range of central MBH masses, accretion rates, sizes, and orientations of the AGN components strongly influences observational manifestations among the AGN population~\citep{2017A&ARv..25....2P,2018ARA&A..56..625H}.

There are a variety of methods used to estimate MBH masses in modern astronomy, such as modeling the velocities of masers \citep{2019ApJ...886L..27R} or optical stars orbiting within the sphere of influence of an MBH \citep{2008ApJ...689.1044G,2010RvMP...82.3121G}, gravitational lensing \citep{2022ApJ...930L..12E}, and reverberation mapping \citep[RM;][]{1973ApL....13..165C,1982ApJ...255..419B,2021iSci...24j2557C}. 
RM has emerged as a crucial technique in dissecting the unresolved components around MBHs, given its ability to measure light travel time delays ($\tau$) between the fluctuations in the AD continuum and the BLR or dust torus responses. Not only can $\tau$  provide a characteristic size ($R =\tau \cdot c$, with $c$ being the speed of light), but under certain conditions, it also offers insights into the geometry of these regions \citep[e.g.,][]{2020ApJ...891...26A,2017ApJ...849..146G}.
Many indirect methods for estimating the mass of MBHs are also used in the community (see e.g., \cite{2013ApJ...775..116R} for BH mass estimation using broad emission lines, \cite{2015ApJ...813...82R} for $M_{BH}-M_{*}$ scaling relation between the BH mass and stellar mass of the galaxy, \cite{2019ApJ...871...80G} for the fundamental plane of BH activity, and \cite{2024MNRAS.528.3417G} for comparisons of indirect approaches). 
In this work, we consider (a) the so-called single-epoch (SE) MBH mass estimator 
$M_{BH}(\text{FWHM}_{broad~line}, L_{broad~line})$, where the broad line luminosity is a proxy for the BLR radius using $R_{BLR}-L$ relations~\citep{2013ApJ...767..149B,2023ApJ...953..142C,2024ApJS..275...13W}, calibrated by RM. The broad line full width at half maximum (FWHM) tracks the average velocity of BLR clouds under the assumption of virialized motions due to the dominant gravitational influence of the central MBH; (b) the empirical $M_{BH}-\sigma_{*}$ scaling relation between stellar velocity dispersion and MBH mass \citep[e.g.,][]{2000ApJ...539L...9F,2000ApJ...539L..13G}, usually explained as the result of the co-evolutionary growth of MBHs and spheroid stellar mass. 

Compared to supermassive black holes (SMBHs; $> 2\times10^{5}  M_{\odot}$), our knowledge of intermediate-mass black holes (IMBHs; $\leq 2\times10^{5} M_{\odot}$) and lower-mass SMBHs ($2\times10^{5} M_{\odot} < M_{\rm BH} \leq ~2\times10^{6} M_{\odot}$) is still far more limited. This is mainly due to their smaller sizes and weaker emission lines ($\sim1\%$ of the total integrated flux in a given band),  making it difficult to measure the time delays between the AD and BLR, ranging 
from minutes to days for AD-to-BLR delays. The only known example of such measurements is a nearby galaxy NGC~4395 (H$\alpha \sim 7\%$ contribution in the $R$-band), where AD-to-BLR delays are consistent with a scaling relation of $\tau \propto L_{5100}^{0.5}$ \citep{2012ApJ...756...73E,2023AAS...24133606B}. NGC~4395 is located in a relatively unexplored region near the low end of the $R_{BLR}-L$ relation; namely, $M_{BH} = 3.6\pm1.1\times 10^{5} M_{\odot}$ from the first BLR RM of this source \citep{2005ApJ...632..799P} using CIV light echo; $M_{BH} = {4}_{-3}^{+8}\times 10^{5} M_{\odot}$ from warm molecular hydrogen gas dynamical modeling \citep{2015ApJ...809..101D}; $M_{BH} = 9.1^{+1.5}_{-1.6} \times 10^{3} M_{\odot}$ from BLR RM using H$\alpha$ photometric monitoring \citep{2019NatAs...3..755W}. 
NGC~4395 is bulgeless type 1 Seyfert galaxy \citep{2003ApJ...588L..13F} with signatures of a bar~\citep{2015ApJ...809..101D}. Following a number of works, NGC~4395 appears to be an outlier in the $M_{BH}-M_{*,~sph}$ relation~\citep{2003ApJ...588L..13F,2015MNRAS.449.1526L}, while other works generally remain consistent with the $M_{BH}-\sigma_{*}$ scaling relation~\citep{2019NatAs...3..755W,2021ApJ...921...98C}.
Another   active IMBH candidate with some RM estimates is POX52~\citep{2004ApJ...607...90B}, for which a tentative AD-to-torus delay was detected~\citep{2025ApJ...989L..26R}. The upper limit for AD-to-BLR delay was inferred for SDSS J114008.71+030711.4 (or GH08,~\citet{2004ApJ...610..722G}) active IMBH candidate~\citep{2011ApJ...741...66R}.
%\LEt{ Please avoid two parentheses next to each other by merging the contents of them into a single set.***}

One remarkable feature of AGN is their multiwavelength stochastic variability \citep{1997ARA&A..35..445U}, which is, in some cases, prominent compared to the relatively constant host-galaxy component.
\cite{2021Sci...373..789B} argued for an empirical correlation between $M_{BH}$ and the characteristic timescale at which the variability power spectrum flattens, which could provide an additional indirect method for MBH mass measurements. 
\citet{2025ApJ...980..215K} highlighted the potential of variability selection to find lower mass BHs and lower luminosity AGNs than currently possible with optical spectroscopy for dwarf galaxies; however, there are some potential caveats \citep[e.g.,][]{2025A&A...694A.127B}.

%-------------------------------------------------------------
%                      A rotated One column Table in landscape  
%-------------------------------------------------------------

%
%-------------------------------------------------------------
%                              Table longer than a single page  
%-------------------------------------------------------------
% All long tables will be placed automatically at the end of the document
%

%
%-------------------------------------------------------------
%                              Table longer than a single page
%                                            and in landscape, 
%                    in the preamble, use: \usepackage{lscape}
%-------------------------------------------------------------

% All long tables will be placed automatically at the end of the document
%

%-------------------------------------------------------------
%               Appendices have to be placed at the end, after
%                                        \end{thebibliography}
%-------------------------------------------------------------

The time lag between variations in the AD continuum and the BLR can be a powerful tool for BH mass estimations \citep{2021iSci...24j2557C} when inferring the BLR velocities from spectra. The community widely uses empirical power-law correlations ($R_{\text{H}\beta,~BLR}-L_{5100}$, $M_{BH}-\sigma_{*}$, etc.) for MBH mass measurements \citep[e.g.,][]{2020ARA&A..58..257G}. However, applying these empirical relations to IMBHs requires large (and thus uncertain) extrapolations from the better-established SMBH regime. Establishing calibration objects in the IMBH regime via RM techniques would help reduce or at least mitigate the current uncertainties. 

We expect more compact structures surrounding IMBHs, leading to shorter observed time delays between the AD continuum and BLR emission line. In this way, IMBHs can act ``time-squeezing'' machines and they are especially potent for relatively short monitoring campaigns.
We keep in mind the fact that AGN broad-line variability can also originate from illumination of BLR different radii with different kinematics, which is usually referred to as `BLR breathing' phenomena~\citep{2020ApJ...903...51W}. However, there are cases when BLR breathing cannot explain the change of the delay during campaign, for instance, due to orbital rotation of asymmetrical BLR~\citep[see e.g.][]{2024ApJS..272...26S,2025arXiv250213202G}.

In this work, we refer to highly accreting or high-Eddington-rate or rapidly accreting MBHs, which have a bolometric luminosity higher than 10\% of the Eddington luminosity, $L_{bol}>0.1\times L_{Edd}$. The uncertainties on the accretion rate stand in the way of confidently excluding a possible super-Eddington regime ($L_{bol} > L_{Edd}$) for such sources, where the sphere of BH influence is not resolved. The super-Eddington MBHs usually exhibit smaller time delays than expected from the R-L relations for their luminosities~\citep{2015ApJ...806...22D,2018ApJ...856....6D}.

To check the reliability of the empirical relations for BHs in the IMBH regime, we measured the BLR time lag of a rapidly accreting IMBH and estimated its BH mass. 
AGNs close to the Eddington limit are extreme cases in terms of their high luminosity; thus, they can be detected more easily than their slowly accreting counterparts (a rare case where the Malmquist bias is useful). 
%We focus our first IMBH BLR reverberation mapping campaign on rapidly accreting IMBH candidates.

With this work, we aim to initiate a series focused on high-accretion-rate IMBH confirmation using BLR, AD, and dust RM, which will be strengthened by future ground-based instrumentation mounted at the Extremely Large Telescope (ELT\footnote{\url{https://elt.eso.org/}}). 
The paper is organized as follows. In Section~\ref{sec: obj_ident}, we
discuss the process behind the selection of our target. In Section~\ref{sec: observations}, we describe all the observational data collected for the analysis.
In Section~\ref{sec: rc600_campaign}, we focus specifically on the RM campaign and intra-night variability. In Section~\ref{sec: BH_mass_est}, we describe all the methods we used for the BH mass estimations in this work. In Section~\ref{sec: accretion_rates}, we aim to constrain the BH accretion rate. In Section~\ref{sec: discussion}, we discuss the validity of our findings and potential caveats. Finally, in Section~\ref{sec: conclusion}, we summarize our conclusions.
\section{Object identification and selection}
\label{sec: obj_ident}
The parent sample of  IMBH candidates was obtained using the \href{http://rcsed.sai.msu.ru}{RCSED}\footnote{\url{http://rcsed.sai.msu.ru}}~catalog~\citep{2017ApJS..228...14C}, which compiles spectral energy distributions and spectral fitting results of 800,000 galaxies, with source spectra taken from SDSS DR16 \citep{2020ApJS..249....3A}. The $\sim$1900 candidate IMBHs and low-mass AGN were selected using the broad component of $H\alpha$-line, similarly to the prior study by \citet{2018ApJ...863....1C}. The variability properties of these IMBH candidates are demonstrated on the dedicated webpage\footnote{\url{http://lc-dev.voxastro.org/valc.html}}, which is based on post-processing of ZTF forced photometry light curves~\citep{2022aems.conf..359D,2024ASPC..535..283D}. We carried out a wide follow-up campaign, including Chandra, XMM-Newton, and Swift X-ray observations~\citep{2022aems.conf..304T}. 
We found more than ten objects with an accretion rate around the Eddington limit or even higher (i.e., $\geq50\%$ considering bolometric luminosities derived from the X-ray luminosities adopting the bolometric correction formalism of \citealp{2019MNRAS.488.5185N}). Taking observational constraints and ZTF Forced photometry variability properties (reduced chi-square $\chi^{2}_{ZTF\_g} > 2~\&~\chi^{2}_{ZTF\_r} > 2$) into account, we chose \object{2MASS J14485011+1608030} (also known as \object{LEDA 1501539} and \object{SDSS J144850.08+160803.1}; a Seyfert 1 IMBH candidate~\citep{ 2018ApJ...863....1C, 2018ApJS..235...40L} with a spectroscopic redshift of $z_{\rm spec}=0.0383$~\citep{2015ApJS..219...12A}) as the first rapidly accreting IMBH candidate for our photometric RM campaign using narrow H$\alpha$ filters. We use the abbreviation J1448+16 for this object hereafter.

\section{New and archival observations}
\label{sec: observations}
Table~\ref{tab:all_archival_data_with_MJDs} lists all the archival and new data for J1448+16 analyzed in this work. The Modified Julian Date (MJD) is used for the x-axis of the light curves and is specified in the Table~\ref{tab:all_archival_data_with_MJDs}. Our dedicated follow-up programs in X-ray, narrow H$\alpha$, and SDSS~g' photometric monitoring are highlighted in bold. In the following sections, we discuss the data quality, data reduction, and data
analysis.
\begin{table*}
\centering
\caption{Available data for J1448+16 analyzed in this work. \label{tab:all_archival_data_with_MJDs}}
\begin{tabular}{cccc}
\smallskip
Source & Type & MJD & Date (YYYY-MM-DD)\\
\hline
\hline
{\bf XMM-Newton EPIC (our follow-up)} & X-ray spectrum & 58743 & 2019-09-17\\
SDSS DR16 & 1D optical spectrum (3600--7600\AA) & 54507 & 2008-02-11\\
{\bf RC600 (our follow-up)}& optical light curve in $\{$SDSS~g', H$\alpha \}$& 60383 - 60513 & 2024-03-14/2024-07-22\\
\hline
\end{tabular}
\end{table*}

\subsection{New X-ray XMM-Newton observations}
\label{sec: xmm}
\begin{figure}
\includegraphics[width=\linewidth]{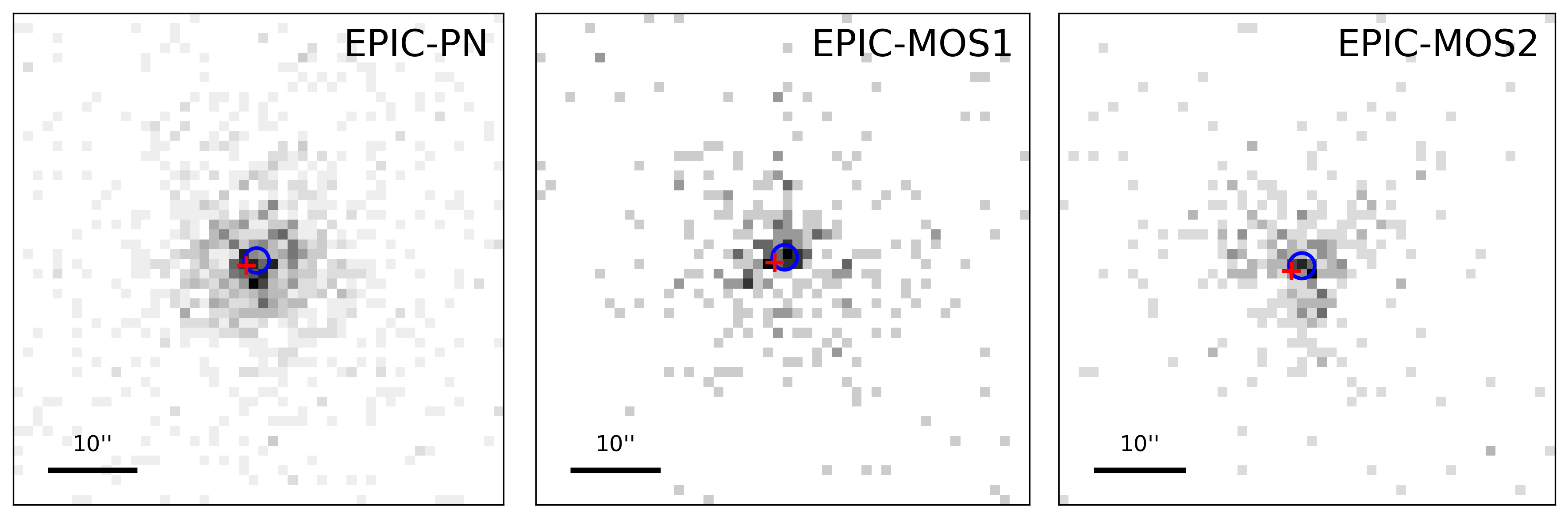}
\caption{X-ray images of J1448+16 from EPIC-PN, EPIC-MOS1, and EPIC-MOS2 detectors in total XMM-Newton energy band (0.2--12~keV). The red cross marks the galaxy center from optical observations, and the blue circle marks the XMM detection center, with the circle size representing the coordinate estimation $3\sigma$ error.  \label{fig:xmm_img}}
\end{figure}

We observed J1448+16 with XMM-Newton EPIC on 07/01/2020 (ObsID 0843440801, PI: Chilingarian), with a total exposure time of 17,700 seconds. We followed the data reduction procedures of the XMM–Newton Science Analysis System (XMM SAS, \citealt{2004ASPC..314..759G}) to extract the calibrated event lists from the observational data. In the calibrated event lists, J1448+16 shows a clear detection of the X-ray signal at coordinates RA = 14:48:50.01 and Dec = +16:08:03.74, which matches the optical target within $3\sigma$ of the XMM spatial uncertainty of 1.38 arcsec (see Figure \ref{fig:xmm_img}). After extracting the spectra from the calibrated exposures of the EPIC-pn, EPIC-MOS1, and EPIC-MOS2 detectors, we combined the spectra from different detectors to achieve a higher signal-to-noise ratio (S/N), obtaining a total of 1,000 background-subtracted photons in the 0.2--10~keV energy range. For the spectral extraction, we used circular regions with radii of 10'' and 90'' for the source and background, respectively. The center of the source region was placed at the X-ray detection coordinates, while the background regions were selected to avoid detector gaps and background point sources.

\begin{figure}
\includegraphics[width=1\linewidth]{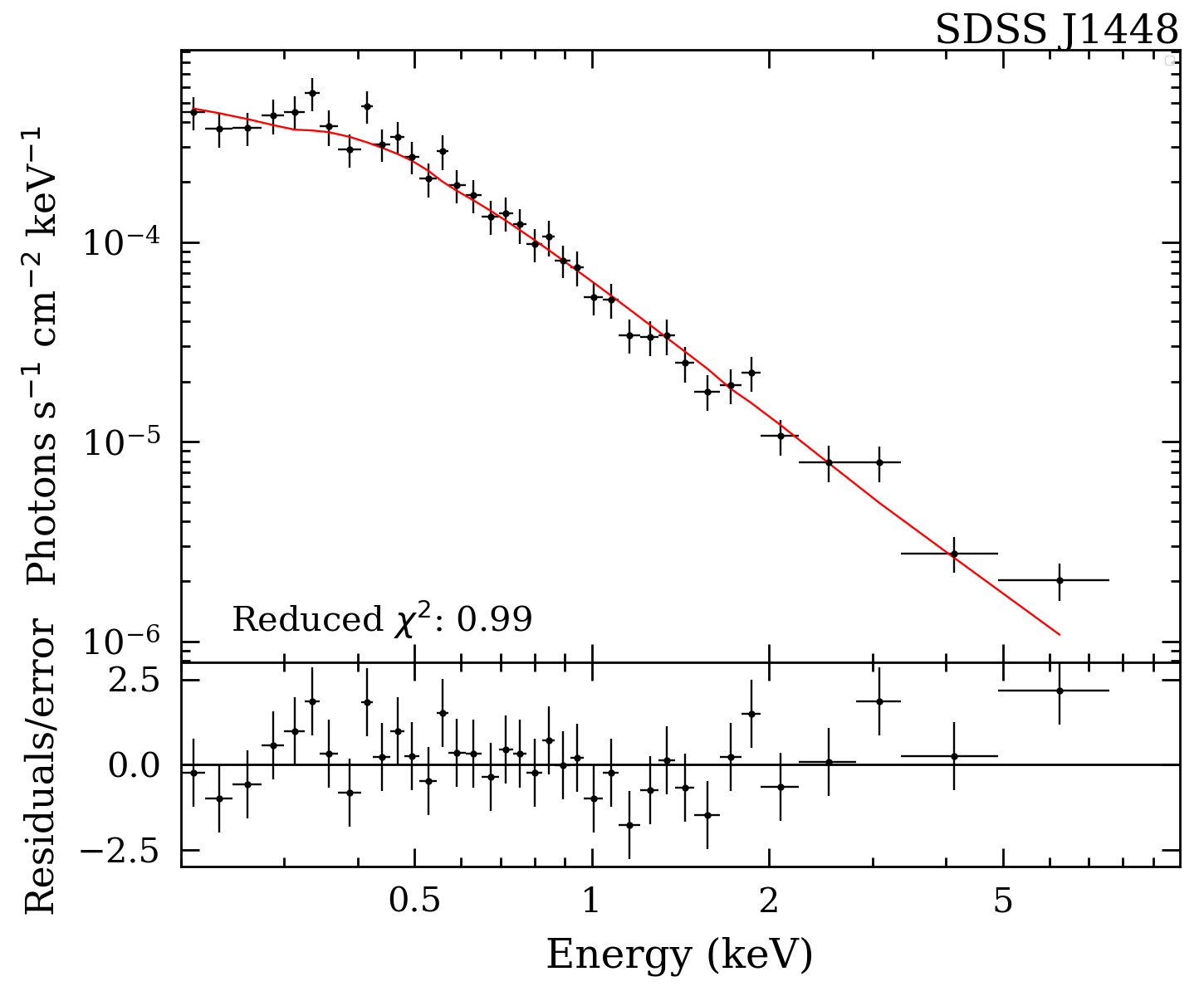}
\caption{X-ray spectrum of J1448+16 from combined EPIC-PN, EPIC-MOS1, and EPIC-MOS2 exposures. The solid red line represents the best-fit model, which includes a power law with photoelectric absorption with photon index of $\Gamma = 2.32^{+0.15}_{-0.13}$ and hydrogen column density of $\mathrm{nH} = 9.7^{+11.4}_{-9.2}\times10^{19}\mathrm{\;atoms/cm}^2$. \label{fig:xmm_spec}}
\end{figure}

To model the X-ray spectrum, we used the Python package {\tt\string Sherpa} \citep{2023zndo...7948720B}, which implements XSPEC \citep{1996ASPC..101...17A} models: a simple photon power-law component to describe AGN activity and a photoelectric absorption component to account for line-of-sight absorption. Sherpa includes a built-in background subtraction procedure. The best-fitting model (see Figure \ref{fig:xmm_spec}) yields a power-law photon index of $\Gamma = 2.32^{+0.15}_{-0.13}$ and a hydrogen column density of $\mathrm{nH} = 9.7^{+11.4}_{-9.2} \times 10^{19}\,\mathrm{atoms/cm}^2$, with errors corresponding to the 90\% confidence level. The low hydrogen column density confirms J1448+16 as an unobscured AGN. A photon index of $\Gamma \geq 2$ is indicative of weak Comptonization and a soft state%, characterized by a strong disk and a soft tail 
~\citep{2007A&ARv..15....1D}, and generally is associated with a high accretion rate~\citep{2013MNRAS.433.2485B}.
\citet{2013MNRAS.433.2485B} reveals for the AGN sample the relation between photon index $\Gamma$ and Eddington rate $\epsilon$, where the inference corresponds to the super-Eddington regime for J1448+16's photon index. The X-ray luminosity in the 0.2--10 keV energy range, estimated from the model spectrum, is $1.17^{+0.16}_{-0.14} \times 10^{42} \mathrm{erg/s}$, which agrees with the luminosity from the 4XMM-DR14 catalog \citep{2020A&A...641A.136W}.

\subsection{Archival optical SDSS spectrum}
\label{subsec:sdss_spec}

J1448+16 was observed in the SDSS survey \citep{2000AJ....120.1579Y} and its spectrum is publicly available. The SDSS spectrum has been automatically analyzed within the \href{http://rcsed.sai.msu.ru}{RCSED} project, using the {\sc NBursts} \citep{2007MNRAS.376.1033C,2007IAUS..241..175C,2012IAUS..284...69K} full spectrum fitting technique. As part of the upcoming \href{https://rcsed2.voxastro.org/}{RCSEDv2}\footnote{\url{https://rcsed2.voxastro.org/}} release~\citep{2024ASPC..535..371R,2024ASPC..535..175G,2024ASPC..535..243K,2024ASPC..535..179C,2024ASPC..535..375K,2024ASPC..535..415T}, it has been reanalyzed with an updated {\sc auto-NBursts} method (\citealp{2024IAUGA..32P2683R}; Rubtsov,~in prep.), which simultaneously models both stellar and multi-component gas contributions to the spectrum.

We detected 32 emission lines in the full spectral range ($3810 - 9185 \AA$), separated into two kinematic groups: 26 NLR lines (described by a single Gauss-Hermite profile) and 6 BLR lines of the hydrogen Balmer series. The latter group is defined by (i) a Lorentzian profile with broad wings for the case of a spherical distribution of BLR clouds; and (ii) a Gaussian profile for the case of disky distribution of BLR clouds;  \citealp{2013A&A...549A.100K}). We notice that after the convolution with the line spread function (LSF), the Lorentzian profile effectively changes to the Voigt profile. The relatively large LSF (due to the low spectral resolution of SDSS) results in a small contribution of the Lorentzian part, causing degeneracy in the calculation of the distribution parameters. Therefore, for BH mass estimation, we use only the Gaussian BLR profile. The line-of-sight velocity distribution (LOSVD) is described by a Gauss-Hermite function with non-zero higher kinematic moments ($h_3$ and $h_4$ coefficients) for the NLR. We also used additive and multiplicative continua in the form of low-order Legendre polynomials to account for the AGN continuum and extinction. For the J1448+16 fit, we used an additive continuum degree~=~3 and multiplicative continuum degree~=~15. To model the stellar continuum, we employed several available and widely used simple stellar population (SSP) models: PEGASE (\citealt{2004A&A...425..881L}, $R\sim10000$), MILES (\citealt{2010MNRAS.404.1639V}, $\mathrm{FWHM} = 2.3 \AA$), E-MILES (\citealt{2016MNRAS.463.3409V}, $\mathrm{FWHM} = 2.5 \AA$ in the SDSS wavelength range), and X-Shooter (\citealt{2022A&A...661A..50V}, $R\sim10000$). The additive continuum corresponds to the AGN continuum and potential imperfections of sky background subtraction. The additive continuum at the rest-frame $5100~\AA$ $L_{5100}$ is often used as a proxy of the bolometric luminosity~\citep{2019MNRAS.488.5185N}. However, in this work, we used $L_{X}$ and $L_{H\alpha}$ as proxies for the continuum luminosity due to possible degeneracy between the SSP and additive continuum caused by the small number and inherent weakness of the observed absorption lines.

\begin{figure*}
    \centering
    \includegraphics[width=0.98\linewidth]{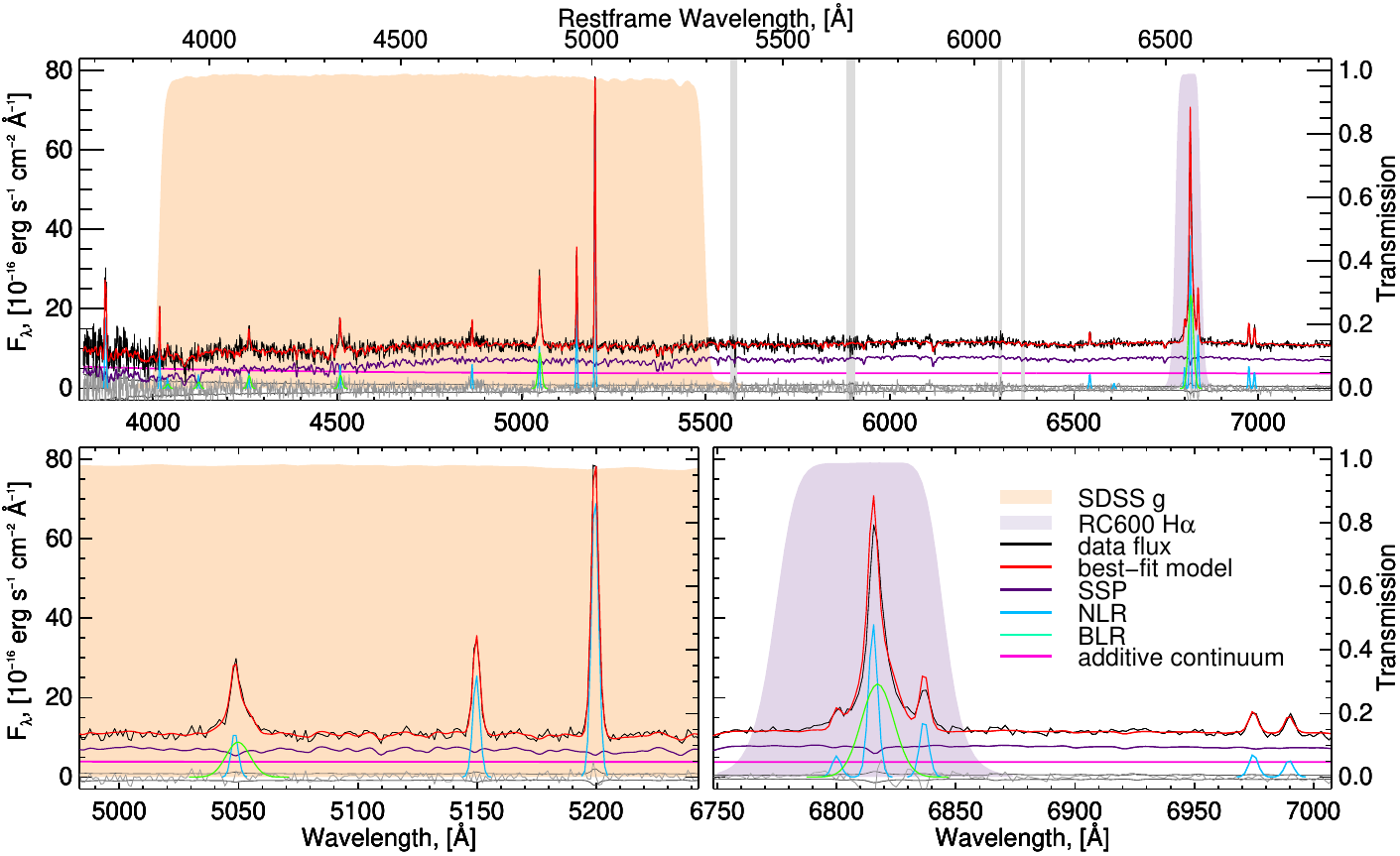}
    \caption{
    {\sc NBursts} fitting results for SDSS spectrum J1448+16 using X-Shooter SSP models and Gaussian BLR profiles. Upper ticks correspond to rest-frame wavelengths, while the middle and bottom ticks correspond to observed wavelengths. {\bf Top panel:} Full wavelength range of the SDSS spectrum; {\bf Bottom panels:} Zoom-in on the surrounding ranges of two emission lines $\mathrm{H\beta}$ and $\mathrm{H\alpha}$. In all panels: the black line corresponds to spectrum fluxes; the red line to the best-fit model; the purple line to the stellar population model including multiplicative continuum (stellar component); the light pink line to the additive continuum component describing the AGN continuum; the light blue and green lines to emission lines templates for NLR and BLR, respectively; the dark gray line to the data-model residuals; the colored shaded regions to the filter transmission curves of SDSS~g\'~in light orange and narrow $\mathrm{H\alpha}$ in light purple; and, finally, the light gray shaded bands to masked and excluded regions from the fit.
    \label{fig:spec_nbursts}}
\end{figure*}

The properties of the stellar populations (age and metallicity) for different SSP models, the velocity dispersions of the NLR and BLR emission components, the fluxes of the two emission lines ($\mathrm{[OIII]}$ and $\mathrm{H\alpha}$), and BH mass estimates are presented in Table~\ref{tab:spec_nbursts}. According to the BPT classification scheme of \citet{kewley06}, the galaxy is located in the AGN region. We illustrate the results of full-spectrum fitting in Figure~\ref{fig:spec_nbursts}.

\section{Broad line region reverberation campaign}
\label{sec: rc600_campaign}
Conducting the BLR RM requires observations in at least two filters to measure the delay between the AD emission fluctuations and the BLR response. 
An efficient and accurate BLR RM also requires the availability of many non-variable calibration stars in the FoV, which therefore benefits from observations over a wide field of view (FoV). 

The integration time for individual frames must be smaller than the expected variability timescales of the target and not lead to the saturation of the calibration stars. In the case of small telescopes and distant objects, the photometric spatial decomposition of the host galaxy component and the AGN cannot be performed. However, in our particular case of a dwarf elliptical galaxy, we have a relatively weak host galaxy contribution to the continuum near the $H\alpha$ line, which opens up the opportunity to use a small telescope for monitoring. \citet{2025ApJS..276...48S} advocated for the use of small aperture telescopes with a broadband to track the continuum and a narrowband to track the broad emission line. In brief, this is effectively our setup. In the following subsections, we describe the monitoring design, reduction of the taken photometric data, aperture photometry, and flux calibration using 33 calibration stars in the field.
\subsection{Design of the reverberation mapping campaign}
\label{sec: design_blr_rm}
We carried out SDSS~g' and H$\alpha$ photometry (PI: Demianenko) with the 0.6-m telescope (RC600) of the Caucasus Mountain Observatory (CMO) of Sternberg Astronomical Institute, Moscow State University (SAI MSU) from March through July 2024. The individual fractional component contributions to the total flux of broad SDSS~g' filter are 96\% for the host galaxy + AGN continuum, 1\% for BLR emission, and 3\% for NLR emission; in the narrow H$\alpha$ filter, these contributions are 66\% for the host galaxy + AGN continuum, 20\% for BLR emission, and 14\% for NLR emission. These estimates come from the SDSS spectrum fit using X-Shooter SSP models (see the description of the fit and decomposition in Section~\ref{subsec:sdss_spec}). The telescope is equipped with an Andor iKon-L BV camera (2048×2048 pixels, with a pixel size of 13.5 $\mu$m; see \citealp{2020ARep...64..310B}, for details) and a set of Johnson-Cousins, SDSS, and narrowband filters. To measure the H$\alpha$ line, we used a custom redshifted H$\alpha$ filter suitable for $0.033<z<0.044$ and covering wavelengths from $673-689$~nm. The monitoring was designed to have a cadence of at least $\frac{\tau_{BLR}}{2}$ days in accordance with the Nyquist theorem.

To predict the H$\beta$ BLR radius, we used the scaling relation $R_{H\beta}(L_{5100})$ of \cite{2013ApJ...767..149B}   (Eq.~\ref{eq: hbeta_r_blr_l5100})
and the relation between $L_{5100}$ and $L_{X,~2-10~keV}$ \citep{2013peag.book.....N}, expressed as  
\begin{equation}
   \log_{10}\left(L_{5100}\right) = 1.4 \log_{10}\left(L_{X,~2-10~keV}\right) - 16.8.
    \label{eq: bolometric_cor}
\end{equation}
The X-ray luminosity\footnote{The {\sc sample\_flux} method of {\sc Sherpa} v4.17 was used for the generation of the X-ray flux distribution of a model, adopting sample size of 5001.} is $L_{X,~2-10~keV} = 3.33^{+0.49}_{-0.44}\times10^{41}$erg~s$^{-1}$ following the best-fit model, described in Section~\ref{sec: xmm}. To calculate the H$\alpha$ BLR radius we used Monte Carlo sampling to account for the inner scatter and coefficient uncertainties of aforementioned relations, we took the relation between $\mathrm{R_{BLR,~H\alpha}}$ and $\mathrm{R_{BLR,~H\beta}}$ as 1.68:1 from \citet{2023ApJ...953..142C}:
\begin{equation}
   R_{BLR,~H\alpha} = 1.68\cdot R_{BLR,~H\beta} = 2.16^{+1.34}_{-0.83}~[\mathrm{lt-days}].
\end{equation}

The monitoring spanned 130~days and was scheduled every night with at least three frames taken in each filter (corresponding to 1800s exposure time). The weather success rate of observations is 15.4\% (20 successful nights; see Figure~\ref{fig:rel_var_ampl_whole_lc} for distribution). 
The AGN in the coadded images (binned over each night) is characterized by mean S/N values of $\sim37$ for the g band and $\sim19$ for the $H\alpha$ band, while among all individually observed frames, the mean S/N values are 20 for the g band and 10 for the $H\alpha$ band.

\subsection{Data reduction}
We performed raw data reduction using master bias, dark, and flat frames provided by the observatory using the standard approach: 
\begin{equation}
    \rm{Data} = \frac{\rm{RawData}/\tau_{Data} - \rm{Dark}/\tau_{Dark}}{(\rm{Flat}-\rm{Bias})/\rm{median(Flat)}}.
\end{equation}

\begin{equation}
    \rm{Error} = \frac{\sqrt{\rm{RawData}}/\tau_{Data}}{(\rm{Flat}-\rm{Bias})/\rm{median(Flat)}}
\end{equation}
The read-out and dark current noise are negligible and they were therefore not considered in the full error budget. 
The flat field is relatively stable for the camera. However, we still used three flat frames obtained at different dates for each filter (26 Feb, 22 May, and 01 Jun for g-band and 31 Mar, 21 May, and 01 Jun for H$\alpha$-band), associating them with the observations closest in time. Eight of the frames have satellite trails, and we verified that none of them overlap with the target nor with the stars selected for flux calibration. We discarded one of the 20 epochs (May 22) due to its inconsistency with the corresponding flat field, which resulted in the fluxes of the reference stars being significantly offset compared to the fluxes from all other epochs. Additionally, we noticed a focusing problem in two out of three H$\alpha$ frames (probably caused by manual filter change) taken on April 9, so we removed these frames from consideration since an absolute flux calibration could not be performed. After the image reduction, we subtracted a 2D background model to correct imperfections in the flat field correction: we used the {\sc Photutils} {\sc Background2D} method with $3-\sigma$ clipping, {\sc MedianBackground} as a background estimator, and a $50\times50$ pixel window.

\subsection{Photometric measurements and flux calibration}
\label{sec: flux_calibration}
The photometric measurements of J1448+16 and other sources in the RC600 frames were performed as aperture photometry with a $3''$ diameter aperture (similar to SDSS) using the \texttt{Photutils python} package \citep{larry_bradley_2023_7946442} with a mean background estimated in a $2''$ wide annulus around the aperture.

We calibrated the flux using 33 nonsaturated, non-variable stars from the images, which are shown in Fig.~\ref{fig:foV} by 6'' diameter magenta circles around them with S/N in the annotation. They were selected from a crossmatch between Gaia DR3 \citep{2023A&A...674A...1G} and SDSS DR12 \citep{2015ApJS..219...12A} catalogs in a 12 arcminute cone around the galaxy with the additional conditions of having a total proper motion $< 20 $~mas~yr$^{-1}$, no variable flag ({\sc phot\_variable\_flag}~$\neq$~{\sc VARIABLE}), {\sc classprob\_dsc\_combmod\_star}~>~0.9, and SDSS g magnitude between 15.7 and 18 (to ensure sufficient S/N in the science images and avoid saturation). We also added a subset of 54 fainter stars with $18<g<19.5$ as a validation sample to control our correction algorithms. Validation stars are shown by dark cyan circles with S/N in the annotation in Fig.~\ref{fig:foV}.
The ADU flux of these stars was calibrated to their flux in SDSS filters using the powerful tool for the linear fit of data with significant outliers:\ the random sample consensus (RANSAC; \citealp{10.1145/358669.358692}) algorithm, constituting the absolute flux calibration and accounting for CCD sensitivity and filters throughput.
Additionally, we provided a relative flux calibration by subtracting the median variation of the calibration stars at each epoch in magnitude space from the AGN magnitude. This resulting correction is below the level of flux uncertainties ($\pm1\%$). However, this calibration step is useful for updating the AGN flux uncertainty to be realistic.

\begin{figure*}
\centering
\includegraphics[width=0.99\linewidth]{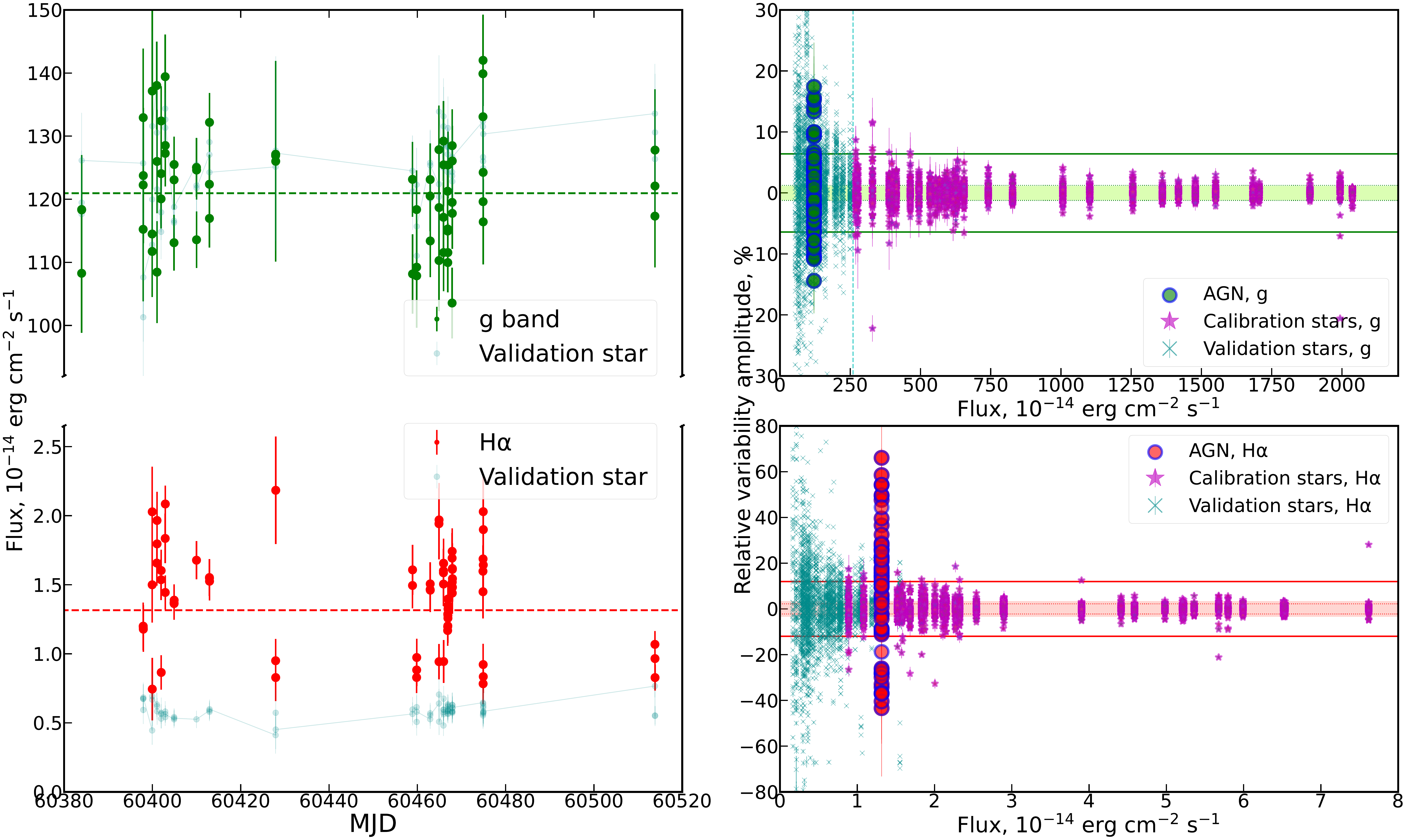}
\caption{Edge colors for markers in the right graphs correspond to the object classes: blue is AGN, magenta is calibration stars, and dark cyan is validation stars. 
{\bf Left:}  RC600 light-curve of J1448+16 binned by individual nights. The green points correspond to SDSS~g' flux, the red points correspond to H$\alpha$ flux.  
The pale cyan line represents the weighted mean validation star. {\bf Right:} Relative variability amplitude in g' and $H\alpha$ bands. The large circles show the variability of the AGN, the magenta stars demonstrate the variability of 33 calibrating stars (due to measurement uncertainties) and the crosses represent the validation sample of 54 faint stars. The vertical cyan line denotes the separating flux (in $g'$-band) between calibration and validation samples. \label{fig:rel_var_ampl_whole_lc}} 
\end{figure*}

\subsection{Variability, normality, and stationarity tests}
\label{sec:variability_metrics}
The light curves in Figure~\ref{fig:rel_var_ampl_whole_lc}~(left) show significant variability established by the use of $\chi^2$ statistical criteria to compare with a constant (we considered two constant models: weighted mean and weighted median if there are outliers) as the null hypothesis. 
We set the threshold for the p-value to $p_{val}<0.05$ to reject the null hypothesis. We are aware of the conceptual p-value threshold problems~\citep{Wasserstein29032019} and consider the p-value suitable for the given number of points and S/N compared to complex DRW models with Bayesian inference.
\citet{2017MNRAS.464..274S} claim the unreliability of the $\chi^2$ test as the variability metric, since the residuals y-$y_{weighted}$/$\sigma$ can produce a non-Gaussian distribution if systematics play a significant role. Therefore, we additionally consider the Anderson-Darling and Kolmogorov-Smirnov statistical tests to compare the residuals with a standard normal distribution (specified as normality tests in the Table~\ref{tab:var_lc}). We confirm the normality of the residuals in SDSS~g\'~,~however, the AGN residuals of H$\alpha$ do not correspond to a standard normal distribution due to a small number of strong ``outliers.'' In the H$\alpha$ case, we trust the weighted (by errors) median as a more robust constant estimator. The variability columns in Table~\ref{tab:var_lc} list the p-values and reduced $\chi^2_{red}$. The $\chi^2$ values confidently indicate the presence of variability in H$\alpha$, SDSS~g\'~passbands.
To check the reliability of the statistical tests, we used the light curve of the weighted mean flux (over 54 faint validation stars) with median uncertainty as a reference
to keep the S/N regime the same as in AGN. The validation star does not show significant variability in both bands, which confirms that AGN variability comes from internal processes. 

%We propose to use the statistical criteria to check the stationarity of light curves, and then apply DRW only for stationary time series. 
DRW is a weakly stationary process, therefore, it requires the mean, variance, and covariance of the random process to be time-invariant. This work uses the augmented Dickey-Fuller (ADF) unit root test for the proposed check of stationarity. The P-values of the null hypothesis are listed in Table \ref{tab:var_lc}, indicating that the whole AGN light curve demonstrates stationarity. Thus, it is justified to apply the DRW model in {\sc JAVELIN} and  
the GPCC time lag extractor described in Section~\ref{subsubsec: GPCC}. However, the intra-night subset does not show significant stationarity to justify DRW modeling. 

\begin{table*}
\begin{threeparttable}
\centering
\caption{Stationarity, variability, and normality properties of investigated light curves. \label{tab:var_lc}}
\begin{tabular}{l|c|cccc|cc}
\hline
\multicolumn{1}{|c}{} &
\multicolumn{1}{|c|}{Stationarity} &
\multicolumn{4}{c|}{Variability} &
\multicolumn{2}{c|}{Normality}\\
\hline
\hline
%\smallskip
Filter & ADF $p_{val}$ & W. mean $p_{val}(\chi^2)$ & W. mean $\chi^2_{red}$ & W. median $p_{val}(\chi^2)$ & W. median $\chi^2_{red}$ & KS $p_{val}$ & AD $p_{val}$\\
\hline
\multicolumn{2}{c}{} \\
\hline
\multicolumn{8}{c}{AGN} \\
\hline
H$\alpha$~(all frames) & $0$ & $0$ & $ 5.23  \pm  0.77$  & $0$ & $7.36 \pm 1.43$ & $0$ & $<1\%$\\ 
g\'~(all frames) & $0.00009$ & $0.0001$ & $1.74\pm0.28$ & $ 0.00006$  & $ 1.80\pm0.29$ & $0.171$ & $ >15\%$ \\
\hline
\hline
H$\alpha$~(2024-06-13)\tnote{1} & $0.11820$ & $0$ & $ 7.09\pm1.25$ & $0$ & $12.99 \pm6.08$ & $0.007$ & $5\%$\\
g\'~(2024-06-13)\tnote{1} & $0.94831$ & $0.0554$ & $1.80\pm0.53 $ & $0.02635$  & $ 2.13\pm0.82$ & $0.748$ & $>15\%$ \\
\hline
\multicolumn{2}{c}{} \\
\hline
\multicolumn{8}{c}{Validation star}\\ 
\hline
H$\alpha$~(all frames) & $0.3534$ & $1$ & $0.37\pm0.08$ & $1$  & $0.37\pm0.08$ & $0.034$ & $ >15\%$ \\ 
g\'~(all frames) & $0.00016$ & $0.4352$ & $ 1.00\pm0.16$ & $0.32832$ & $ 1.05\pm0.18$ & $0.596$ & $ >15\%$\\
\hline
\hline
H$\alpha$~(2024-06-13)\tnote{1} & $0.00057$ & $0.9988$ & $0.10\pm0.03$ & $0.99603$ & $0.14\pm0.04$ & $0.173$ & $>15\%$\\
g\'~(2024-06-13)\tnote{1} & $0.49293$ & $0.9628$ & $0.17\pm0.04$ & $0.91712$  & $ 0.24\pm0.09$ & $0.518$ & $>15\%$ \\
\hline
\end{tabular}
\tablefoot{
ADF refers to the augmented Dickey-Fuller stationary test. KS and AD refer to the Kolmogorov-Smirnov and Anderson-Darling test of normality, respectively. For the null hypothesis to be consistent with a noisy constant, we consider two constant models: weighted (by errors) mean and weighted median. Uncertainties (due to different numbers of epochs) for $\chi^2$ are generated using the bootstrap technique. The second part of the table shows the remaining variability level after flux correction for the weighted mean validation star at the same faint flux regime as the AGN. The p-value threshold to reject the null hypothesis is $p_{val}<0.05$.
}
\begin{tablenotes}
       \item [1] Intra-night variability over $\sim$ 2~hours during the monitoring on 2024-06-13 (MJD = 60474)
\end{tablenotes}
\end{threeparttable}
\end{table*}

\subsection{Intra-night variability: IMBH signature}
\label{sec: intra-night}
Intra-night variability on hourly timescales could provide additional confirmation of IMBH nature \citep{2022AJ....163...73S}. 
In NGC~4395, variability on timescales $\sim$ 20 minutes has been detected ~\citep{2023MNRAS.519.3366M}. The blazar-like amplitude of variability hints at the jet activity even in low-mass radio-quiet (using conventional radio-loudness criteria invented for more massive siblings) AGNs with masses of $\sim10^{6}M_{\odot}$ \citep{2023MNRAS.518L..13G}. However, we know that the observed flux of a jet at the given frequency depends on BH mass ($F_{jet} \sim M^{17/12}$; \citealp{2003MNRAS.343L..59H}). This fact leads us to not consider IMBHs with BH masses $\sim 10^{5}M_{\odot}$ as radio-quiet even in the case of radio non-detection. Such "blazar-like" variability can point to the presence of a jet even in the case of radio non-detections. Therefore, it is interesting to investigate short timescale variability, where high amplitudes can hint at a jet presence. %, for the first time for the rapidly accreting IMBH.

\begin{figure*}
\includegraphics[width=0.49\linewidth]{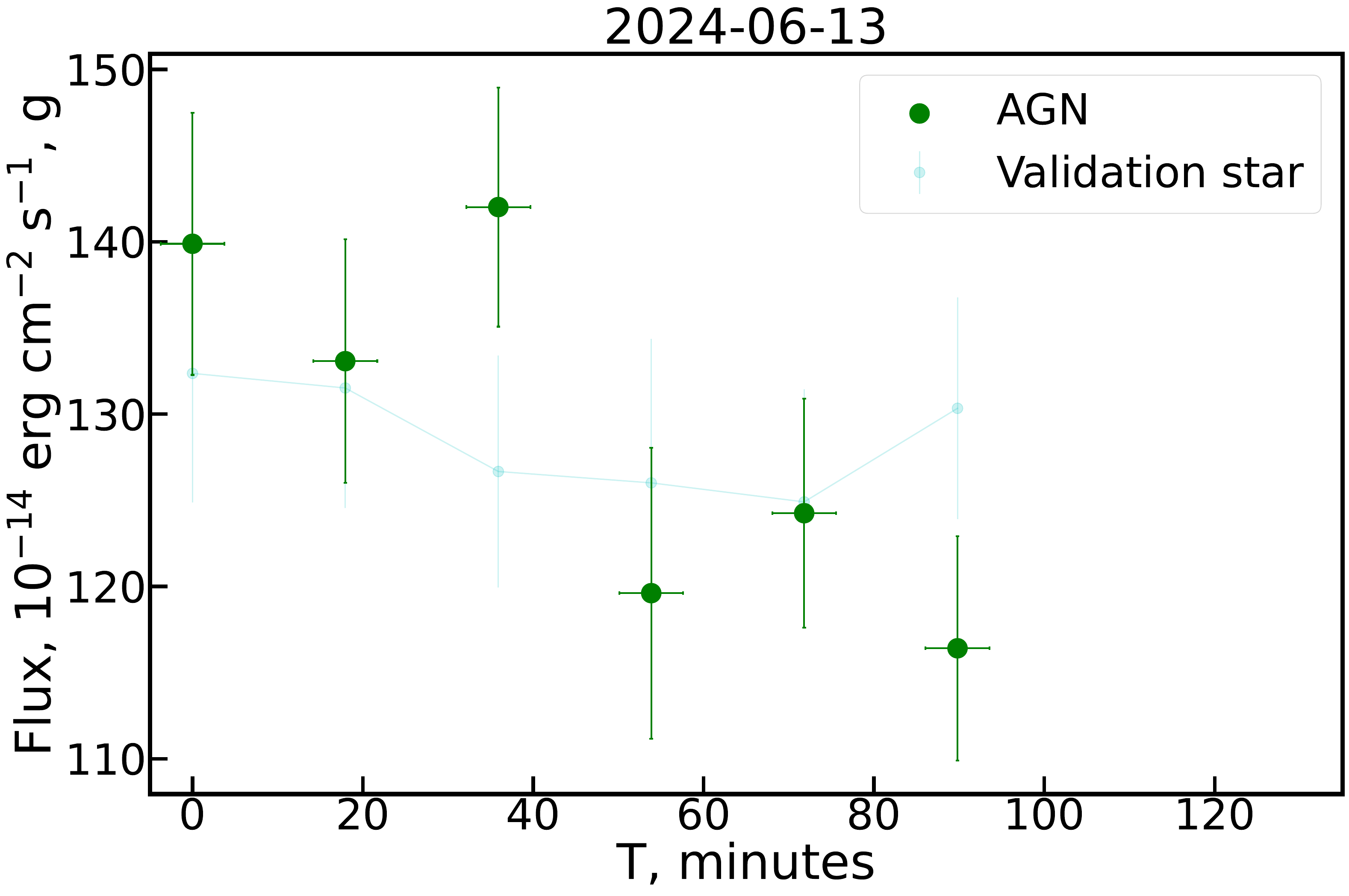}
\includegraphics[width=0.49\linewidth]{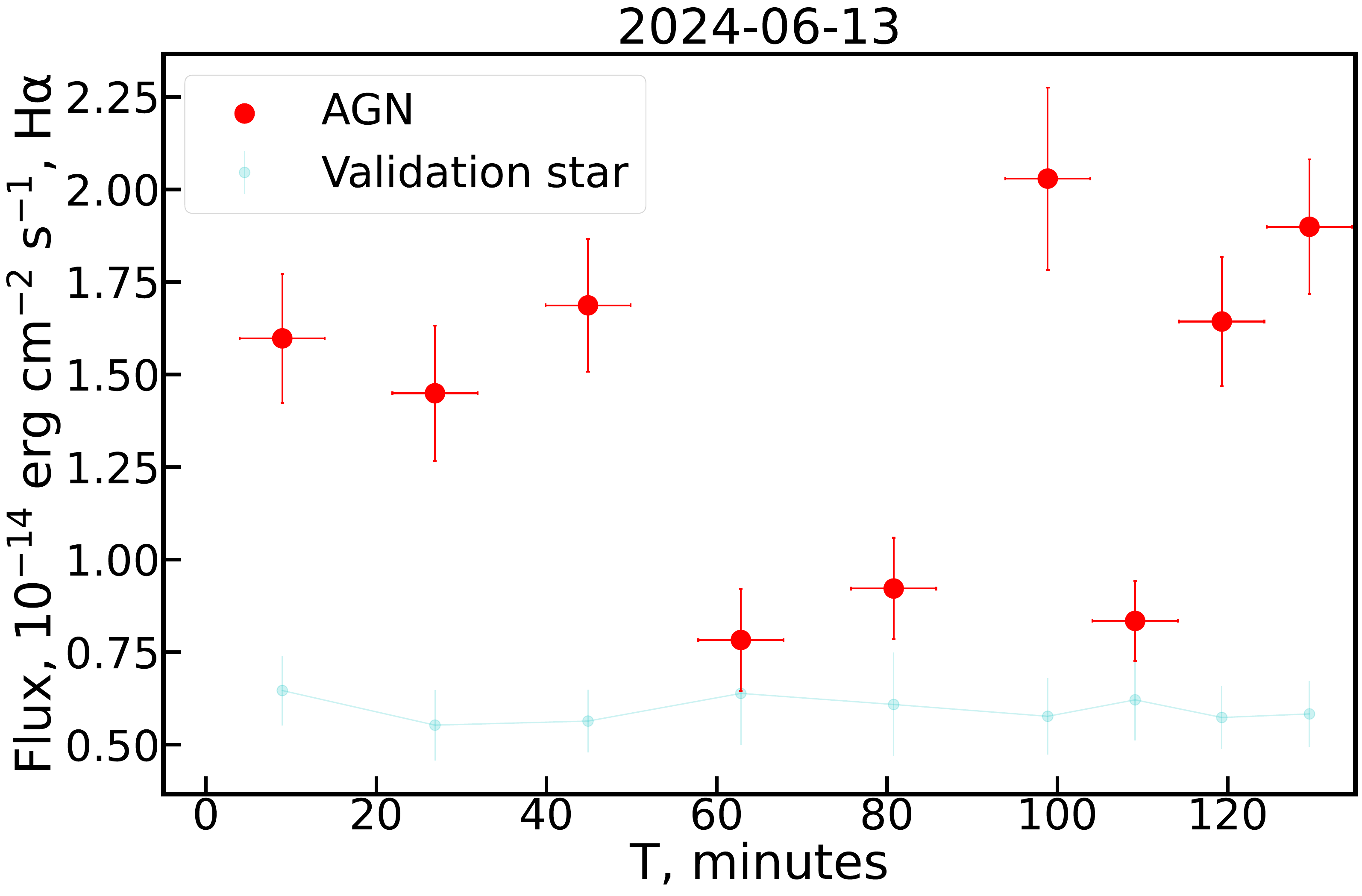}

\caption{
{\bf Left:}  Corrected AGN luminosity in individual g-band frames during the night of 2024-06-13.  
{\bf Right:} Same as left panel, but for H$\alpha$. %\LEt{ Please review as I\ wasn't sure what the "same"\ was referring to.***}
The observations span a time range of $\sim2$ hours, the timestamps are relative to the start of the first g-band exposure. The X-axis error bars represent individual exposure times. \label{fig:rel_ampl_var_intra_night}  }

\end{figure*}

We detected intra-night variability in the H$\alpha$ passband on 13 June 2024, when we had nine H$\alpha$ and 6 $g$\'~band images taken over 2 hours. The relative variability amplitude on the y-axis of Figure~\ref{fig:rel_ampl_var_intra_night} shows the credible detection compared to the residual variability level of calibrating stars (small circles) and validation stars (small crosses) after flux correction. We have claimed the first detection of intra-night variability (on $\sim$1.7~hour timescales) in a high-accreting IMBH, at a level of $\sim55\%$ in narrow H$\alpha$ band and $\sim10\%$ in SDSS~g\'~. The previous study~of~\cite{2024ApJ...974..288Z} found no detectable optical variability for another high-Eddington-rate IMBH candidate, J0249$-$0815, in the AD continuum RM campaign. J1448+16 is hosted in an elliptical galaxy with a stellar mass roughly two times smaller compared to the spiral host of J0249$-$0815. The contribution of sellar light to the broadband is lower, thereby allowing us to observe its AGN variability more prominently on top of a weaker host galaxy continuum. Therefore, it hints at the need to focus optical broadband RM efforts on active IMBHs living in dwarf elliptical hosts. Moreover, the high-amplitude, blazar-like intra-night variability could be caused by a jet~\citep{2023MNRAS.518L..13G} in J1448+16. This assumption can be tested in future studies using high-cadence observations.

We applied the variability metrics from Section~\ref{sec:variability_metrics} to the intra-night light curve. J1448+16 is significantly variable over these 2 hours, considering the p-values of the $\chi^{2}$-test with null hypothesis of a constant with white noise (see  columns 2-5 and rows 5-6 in Table~\ref{tab:var_lc}). The faint validation star does not show significant variability during these 2 hours, again confirming that the AGN variability is internal. The KS and AD normality tests contradict each other, probably due to the small number of points. However, they agree on the normality of AGN H$\alpha$ deviations from the weighted mean, which confirms the reliability of $\chi^{2}$-test results.  

The ADF stationarity test indicates non-stationarity of AGN intra-night light curve due to the small number of points, such that DRW modeling could not be used for the time lag extraction. The linear Pearson correlation coefficient between SDSS~g\'~and narrow H$\alpha$ fluxes over 80 minutes demonstrates significant correlation for AGN ($r_{g,~H\alpha} = 0.988$, $p_{val} = 0.001$) and no evident correlation for the validation star ($r_{g,~H\alpha} = -0.054$, $p_{val} = 0.932$), as an additional signature of the AGN short-term variability.   

\subsection{Broad line region-continuum reverberation mapping}
\label{sec: time-lag_methods}
To study the RC600 light curve in the narrow H$\alpha$ and SDSS~g\', we were inspired by the numerous AGN time lag analyses \citep{2019AN....340..577Z,2020ApJ...892...93C,2018A&A...620A.137R} to collect and compare the best available techniques. 
The cross-correlation function (CCF) is widely used as a standard method for time lag measurements between AD continuum emission and emission line flux \citep{1999PASP..111.1347W}. However, the CCF technique is sensitive to the cadence and duration of monitoring. CCF benefits from campaign durations much longer than the time scale of the investigated process \citep{1999PASP..111.1347W}. In our case, we aimed to monitor J1448+16 for $\sim60$ times longer than the expected $\tau_{BLR}$.

We estimated the time-lag $\tau$ between narrow H$\alpha$ (consisting of 20\% of the BLR emission) and SDSS~g\'~fluxes using two methods listed in the following subsections: interpolated-CCF (ICCF) in~\ref{subsubsec: CCF}, and Gaussian process cross-correlation (GPCC) in~\ref{subsubsec: GPCC}. 
We also attempted to use Just Another Vehicle for Estimating Lags In Nuclei ({\sc JAVELIN},~\cite{2013ApJ...765..106Z}) and running optimal average ({\sc PyROA}~\cite{2021MNRAS.508.5449D}), but we found that it failed to fit the parameters of DRW automatically (parameters converged to the lines with the slopes); therefore, the estimate ($\sim7$~days) was removed from consideration. 
For validation of our time lag analysis, we use the weighted mean faint validation star to further confirm the validity of the AGN variability inference shown in Table~\ref{tab:var_lc}.

\subsubsection{Linear interpolation of the cross-correlation function}
\label{subsubsec: CCF}
 
\begin{figure}
\centering
\includegraphics[width=1.1\linewidth]{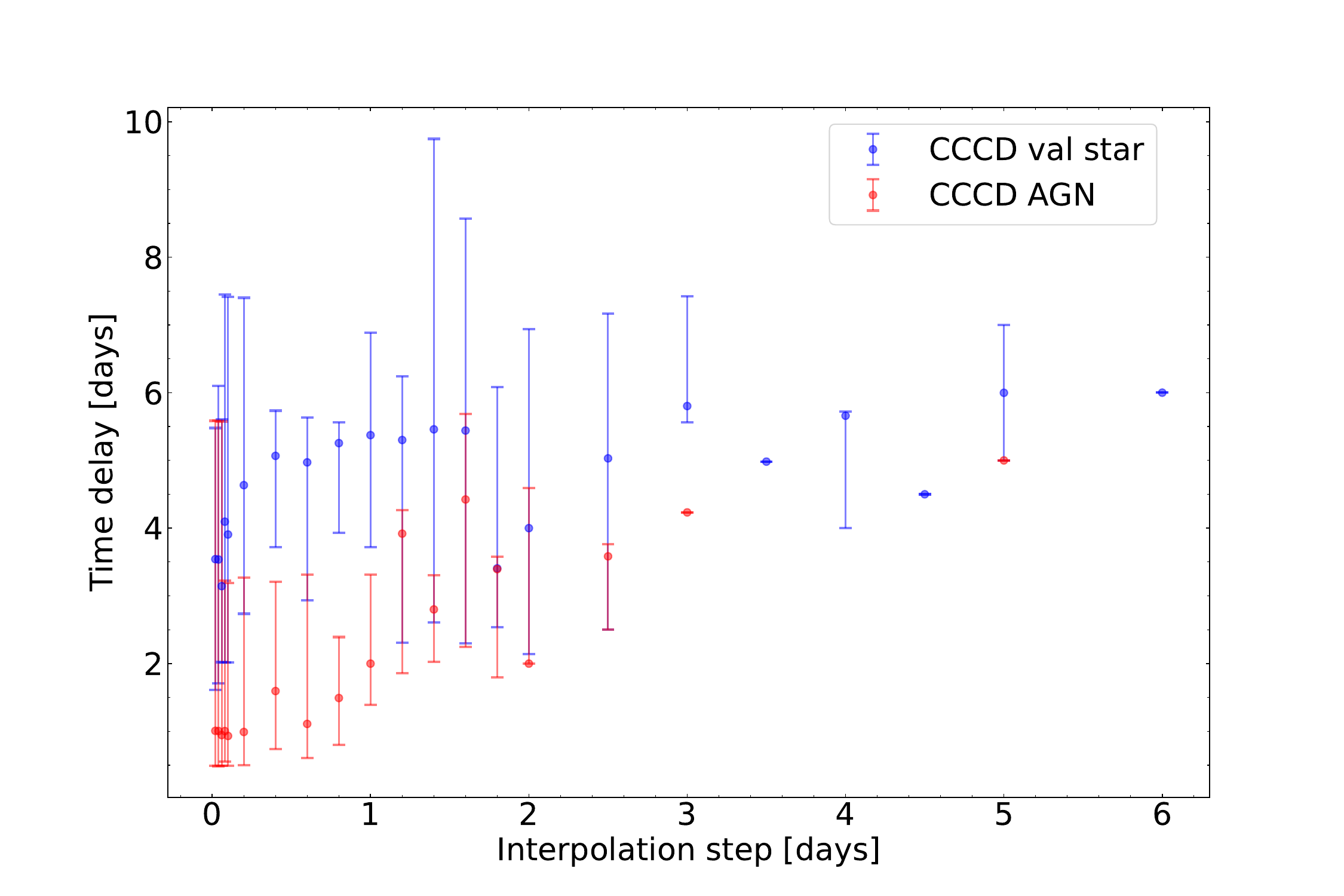}
\caption{Time delay inference dependence on interpolation step of ICCF for whole light curve.  
Color code: blue is the validation star and red is the AGN. Points without error bars correspond to the flat ICCF, where the CCCD failed. The CCCD was run with the following ICCF parameters ({\sc nsim}: 5000, {\sc sigmode}: 0.4, {\sc thres}: 0.6, {\sc nbin}: 100). The Monte Carlo sampling slightly depends on the random seed, but interpolation steps less than one day always show smaller time lags with smaller uncertainties for the AGN compared to the validation star.
\label{fig:iccf_hyperparams_ccpd}} 
\end{figure}

The ICCF \citep{1986ApJ...305..175G,1987ApJS...65....1G} is a linear interpolation between light curve epochs to thin the CCF grid. We used the {\sc pyCCF} Python package \citep{2018ascl.soft05032S} through the {\sc pyPETaL} Python package \citep{2024ApJS..272...26S} to derive a time lag. Different choices of the {\sc interp} hyperparameter for the ICCF, corresponding to the interpolation window, can affect the extracted time lag (see Figure~\ref{fig:iccf_hyperparams_ccpd}).
Thus, we did not rely on a single realization of the ICCF for the time lag extraction. To quantify the uncertainty in the ICCF, we employed a Monte Carlo sampling. Each Monte Carlo realization generates a centroid and a peak of the CCF. By repeating this process many times, we obtained a distribution for the centroids (cross-correlation centroid distribution, CCCD) and peaks (cross-correlation peak distribution, CCPD). We used the CCCD for the time delay inference, which is shown in Figure~\ref{fig:iccf_hyperparams_ccpd}.

The AGN time lag is systematically smaller than the validation star delay  (see~Figure~\ref{fig:iccf_hyperparams_ccpd}) with the tendency to last $\sim1$ day and increase with increasing interpolation steps, as expected. 
The median cadence of the light curve (20 minutes) is much smaller than the mean cadence (1.6 days), which affects the time-lag extraction. To draw a firm conclusion on the validity of the $\sim1$~day time lag, we need to have access to short-cadence observations over an entire\ night for at least one month to maintain stationarity both on small and longer timescales \citep{2021AcA....71..103K}.

\subsubsection{Gaussian process cross-correlation}
\label{subsubsec: GPCC}
GPCC \citep{2023A&A...674A..83P}) is the probabilistic analog of the ICCF. The advantages of this approach include the ability to make an error prediction from the box using Gaussian Processes (GPs) as the model of AGN stochastic variability, along with the use of a prior on the delay.

\begin{figure}
\centering
\includegraphics[width=1.1\linewidth]{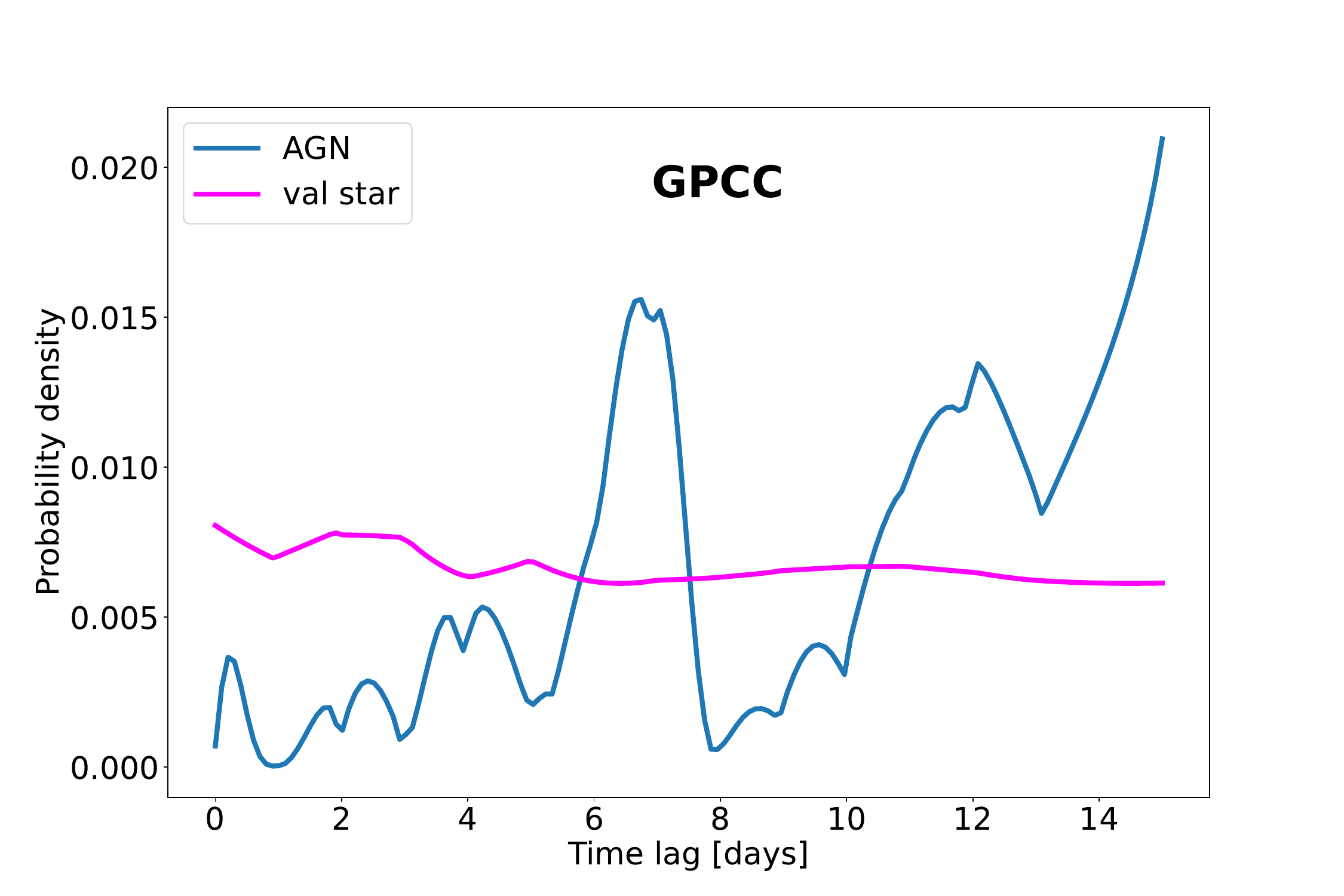}
\caption{GPCC posteriors for the AGN (blue) and the validation star (magenta). GP was run with a uniform 0-15\, day prior for time lag and a Matern1/2 covariance matrix with length scale of l=5.
\label{fig:gpcc}} 
\end{figure}

We used the {\sc Julia} package GPCC~\citep{ascl:2303.006} to derive the time lag, applying a uniform prior of 0 to 15 days. This prior suppresses delays beyond [0,15] by assigning them zero probability. The Matern1/2 kernel (Ornstein–Uhlenbeck, exponential kernel, or damped random walk) was applied as a covariance matrix in the GP with a length-scale hyperparameter equal to five. The length scale corresponds to the smoothing, but is not equal to the interpolation step of the ICCF, discussed in Section~\ref{subsubsec: CCF}.
The GPCC results support the AGN time lag being more prominent than the validation star time lag, which is shown in Figure~\ref{fig:gpcc}. However, the time lag depends on the length scale of the DRW, similar to how the ICCF time lag depends on the interpolation step, and DRW could not be applied on small timescales properly with the current sampling (i.e., non-stationarity is demonstrated for the intra-night variability in Table~\ref{tab:var_lc} with ADF unit root test). Therefore, we could not pin down the $1$~day delay and so, the results were biased toward the  smaller, second ICCF peak, which arises at $\sim7-8$~days. 

\section{Black hole mass estimates}
\label{sec: BH_mass_est}

The basis of empirical relations for BH mass estimations is the virial BH mass, $M_{BH}$, given by the following equation,
\begin{equation}
    M_{BH}=f_\lambda\frac{R_{BLR,\lambda}\cdot\Delta V^{2}_\lambda}{G}
    \label{eq: virial_mass},
\end{equation}
where $\lambda$ signifies a particular emission line chosen for the estimate, $f_{\lambda}$ is the virial factor that can, in principle, depend on wavelength for a disk-like shape of the BLR~\citep{2017FrASS...4...70M}, $G$ is the gravitational constant, $R_{BLR,\lambda}$ is the nominal BLR radius in a selected broad emission line (each line has its own ionization "line-species emission zone", denoted by an average estimated radius, which is Strömgren radius in the first approximation), and $\Delta~V_\lambda$ is the broad emission line width (FWHM). In this work, we fit a single, average FWHM for all broad emission lines (specifically, H$\alpha$ and H$\beta$) due to the assumptions of the {\sc Nbursts} full-spectrum fitting method. The FWHM is measured directly from the spectrum (see Section~\ref{subsec:sdss_spec}).%$-$\ref{subsec:mage_spec})

However, to estimate $R_{BLR,\lambda}$, we require calibrating relations as a link to observable luminosities and there are a few different relations used for this purpose. 
For the BH mass estimation, we used the virial factor $f_{H\alpha} = f_{H\beta} = 1$.
We estimated the BH mass in J1448+16 using FWHM from the SDSS spectrum with different BLR radius estimates, which are described below.

\subsection{Virial calibration in the optical and X-ray }
\label{subsec: MBHs_by_x_ray}
Single-epoch virial calibration using a combination of $R_{\text{H}\beta}-L_{5100}$ \citep{2013ApJ...767..149B}, $L_{\text{H}\alpha}-L_{5100}$ and $\text{FWHM}_{\text{H}\beta} - \text{FWHM}_{\text{H}\alpha}$ \citep{2005ApJ...630..122G} relations has been widely used in the community over the past decade, following, for instance, \citet{2013ApJ...775..116R}. Recently, \citet{2023ApJ...953..142C} derived an independent $R_{\text{H}\alpha}-L_{\text{H}\alpha}$ relation and found that the H$\beta$ BH mass estimator expressed through broad H$\alpha$ as $M_{BH,~\text{H}\beta}(\text{FWHM}_{\text{H}\alpha}, L_{\text{H}\alpha})$ \citep{2013ApJ...775..116R} might be overestimated by up to 0.7~dex at $M_{BH}{<}10^7$~$M_{\odot}$ compared to their $M_{BH,~\text{H}\alpha}(\text{FWHM}_{\text{H}\alpha}, L_{\text{H}\alpha})$ estimates.
For this reason, we adopted the \citet{2023ApJ...953..142C} estimator to calculate the accretion rate in Section~\ref{sec: accretion_rates}. 
Since, in our work, $\text{FWHM}_{\text{H}\beta} \equiv \text{FWHM}_{\text{H}\alpha}$, we did not include a $\text{FWHM}_{\text{H}\beta} - \text{FWHM}_{\text{H}\alpha}$ relation in the analysis, keeping only $R_{\text{H}\beta}-L_{5100}$ from \citet{2013ApJ...767..149B} and $L_{\text{H}\alpha}-L_{5100}$ from \citet{2005ApJ...630..122G}. These relations are listed and explained in detail in Appendix~\ref{sec: appendix_A_single_epoch_mbhs}. We recall that in this work, we have adopted a stratification factor of 1.68 between $R_{\text{H}\alpha}$ and $R_{\text{H}\beta}$~\citep{2023ApJ...953..142C} for the H$\alpha$ BLR radius prediction (described in Section~\ref{sec: rc600_campaign} and given in Table~\ref{tab: bolometric_corrections}). However, we did not use this scaling for the BH mass estimation, since  we can effectively calculate and compare $M_{BH,~\text{H}\beta}$ and $M_{BH,~\text{H}\alpha}$. 

The BH mass estimates based on the single-epoch virial calibration are provided in Tables~\ref{tab: M_BH_spectra} $-$ \ref{tab: MBHs_single_epoch_Gaussian} for different SSP models, resulting in $M_{BH} = 0.9^{+0.9}_{-0.5}-2.4^{+3.7}_{-1.4}~\times10^{5} M_{\odot}$.
Assuming the calibrations are accurate, we might expect to see equality among virial BH mass estimators obtained using different broad emission lines (e.g., $M_{BH,~\text{H}\beta} = M_{BH,~\text{H}\alpha}$). However, this may not always hold true due to unaccounted systematic differences; for instance, between the corresponding virial factors in case of disky BLR geometry (i.e.,  $f_{\text{H}\beta}\sim \text{FWHM}_{\text{H}\beta}^{-1}$ and $f_{\text{H}\alpha}\sim \text{FWHM}_{\text{H}\alpha}^{-1}$ and  $f_{\text{H}\beta}\neq f_{\text{H}\alpha}$~\citep{2017FrASS...4...70M}). 

Given the availability of the X-ray data, it is possible to swap $L_{\text{H}\alpha}-L_{5100}$ for a $L_{5100} - L_{X,~2-10~keV}$ relation (see Appendix~\ref{sec: appendix_B_xray_mbhs}).
We show in Appendix~\ref{sec: appendix_B_xray_mbhs} the BH masses estimated using $L_{X,~2-10~keV}$, the bolometric correction $L_{5100}(L_{X,~2-10~keV})$ \citep{2013peag.book.....N}, a radius following the $R_{BLR,~\text{H}\beta}(L_{5100})$ relation~\citep{2013ApJ...767..149B} or $R_{BLR,~\text{H}\alpha}(L_{5100})$ relation~\citep{2023ApJ...953..142C}, along with the FWHM derived from SDSS full-spectrum fitting using different SSP models. Table~\ref{tab:MBHs_through_LX_gaussian} sum up results for the BH mass 
$M_{BH} = 1.0^{+1.2}_{-0.5}-1.3^{+0.7}_{-0.5}~\times10^{5} M_{\odot}$. To summarize, the X-ray and broad H$\alpha$ BH mass estimates consistently place the source in the IMBH regime, regardless of the chosen empirical relations or SSP models.

\subsection{$M_{\rm BH}$ prediction from the $M_{\rm BH} - \sigma_{*}$ relation}
The SDSS full-spectrum fitting yields a reliable stellar velocity dispersion $\sigma_{*}$ only with the use of the X-Shooter SSP models, since they have the highest resolution in our SSP set.
We show the BH mass predictions from the $M_{BH}-\sigma_{*}$ relation \citep{2000ApJ...539L...9F,2000ApJ...539L..13G} in Appendix~\ref{sec: appendix_C_mbhs_sigma} Table~\ref{tab: mbh_sigma_stellar}. The BH mass expected from the $M_{BH}-\sigma_{*}$ relation~\citep{2013ARA&A..51..511K} is $\sim25$ times larger than the single-epoch BH mass estimates. However, this scaling relation (plus its scatter) should only be treated as an upper limit of $M_{BH}$ \citep{2010ApJ...711L.108B} due to the unresolved sphere of influence of BH for J1448+16. Furthermore, high spatial resolution observations from HST or JWST, for instance, are required for such distances ($D=162.9$~Mpc using $H_{0}=70$~km~s$^{-1}$~Mpc$^{-1}$). 

\section{Constraints on the accretion rate}
\label{sec: accretion_rates}
%In this section, we discuss the accretion rate constraints and the sources of uncertainties in addition to the estimate of BH mass. 
The Eddington luminosity is a fundamental limit given by the equality of radiation and gravitational pressure. For this work, we consider the Eddington luminosity given by the equation from \citet{2013peag.book.....N}: 
\begin{equation}
    L_{\text{Edd}} = \frac{4\pi G M c}{\kappa} \simeq 1.5 \times 10^{38} \frac{M}{M_{\odot}} \, \text{erg} \, \text{s}^{-1},
    \label{eq: l_edd}
\end{equation}
\noindent
where $\kappa$ is the opacity. However, the Eddington luminosity remains an order-of-magnitude estimate, because it implies strict underlying assumptions: the accreting material should be mainly hydrogen (fully ionized) and the accretion should be spherically symmetric with a steady accretion flow \citep{Frank_King_Raine_2002}.
While the fully ionized flow is expected to be seen in AGNs. Realistically, however, the metallicity of the BLR and X-ray corona is rarely zero, which can cause higher opacity $\kappa$ and, hence, a smaller Eddington limit, $L_{\text{Edd}}$. The choice of the bolometric calibration between X-ray luminosity, $L_{X,~2-10~keV}$, and bolometric luminosity, $L_{bol}$, affects the calculation of the accretion rate. In this work, we consider two different approaches for this conversion.

\begin{figure}

\includegraphics[width=\linewidth]{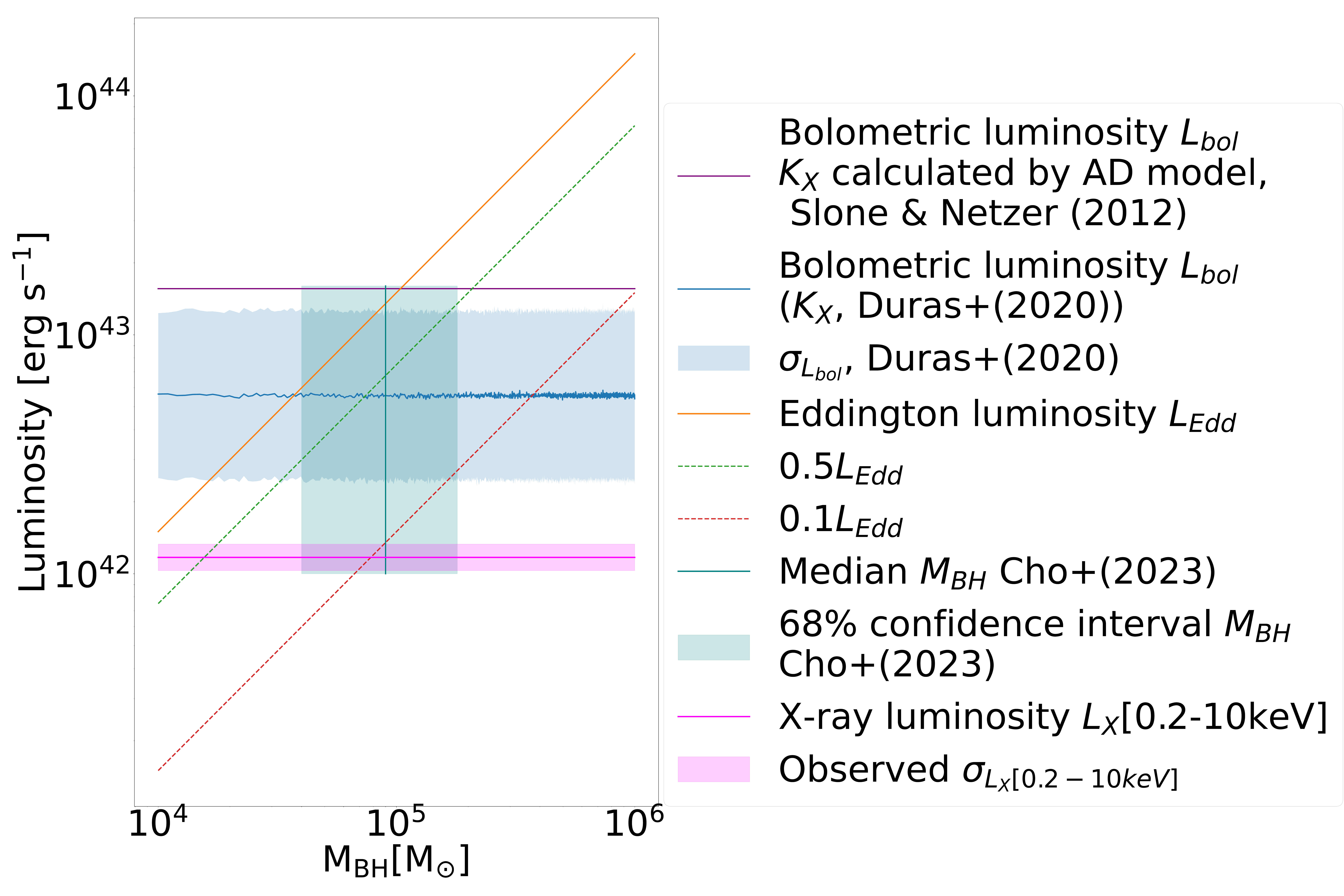}%{}

\caption{Accretion rate constrained by the use of Eddington luminosity and X-ray luminosity. The green vertical line and light green shading denote our single epoch BH mass estimate and 68\% confidence interval using an SDSS spectrum fitted with X-Shooter SSP models and the \citet{2023ApJ...953..142C} $R_{H\alpha}(L_{H\alpha})$ relation. The blue horizontal line and light blue shading correspond to the X-ray derived bolometric luminosity adopting the $M_{BH}$-based bolometric correction equation and 0.35~dex intrinsic 68\% scatter from \citet{2020A&A...636A..73D}. The purple line indicates the simulated bolometric correction derived in this work as the extension of the works by \citet{2019MNRAS.488.5185N} and~\citet{2012MNRAS.426..656S}. Monte Carlo sampling was used for error propagation, incorporating all reported uncertainties on the relation coefficients and inner scatters. In the BH mass interval, the uncertainties of bolometric luminosities are characterized by 68\% confidence intervals for the corresponding values. Luminosities are calculated assuming $z=0.038$, $H_{0}=70$~km~s$^{-1}$~Mpc$^{-1}$.} 
\label{fig: accretion rate}
\end{figure}

\citet{2020A&A...636A..73D} derived the bolometric calibration that depends on BH mass, using the modeling of the observed SEDs of AGN. Following this approach, Fig.~\ref{fig: accretion rate} presents this semi-empirical bolometric luminosity as the blue line with blue uncertainties and Eddington luminosities (Equation~\ref{eq: l_edd}) versus BH mass for different accretion rates using the correction between X-ray and bolometric luminosities, expressed as
\begin{equation}
K_X(M_{\text{BH}}) = (16.75\pm0.71) \cdot \left[ 1 + \left( \frac{\log_{10}(M_{\text{BH}}/M_{\odot})}{(9.22\pm0.08)} \right)^{(26.14\pm3.73)} \right].
\end{equation}
We derived the accretion rate constraint as $L_{bol}/L_{Edd} = 37^{+66}_{-24}-40^{+72}_{-26}$~percent using this bolometric luminosity estimate and accounting for the BH mass constraints estimated from $R_{BLR,~H\alpha}(L_{H\alpha})$. 
The range of accretion rate values corresponds to the range of BH mass derived from the broad H$\alpha$ emission line~\citep{2023ApJ...953..142C} using full-spectrum fitting with different SSP models. Each BH mass has the uncertainty caused by the inner scatter of the R-L relation and measurement errors from the flux and FWHM in the spectrum. This fact leads to individual uncertainties on the corresponding Eddington luminosities and accretion rate estimations. For the uncertainty estimations, we used Monte Carlo sampling and report 68\% of the confidence interval for each accretion rate inference. The last two columns of Table~\ref{fig: r_blr_from_xray} show all individual accretion rate estimates corresponding to different SSP models.

\citet{2019MNRAS.488.5185N} provided a bolometric calibration between $L_X$, $L_{bol}$ and $L_{5100}$ for $M_{BH}\sim10^{7}M_{\odot}$.
The AD model used in \citet{2019MNRAS.488.5185N} adopts the classical thin disk model of \citet{1973A&A....24..337S} and additionally includes full relativistic corrections and Comptonization in the disk atmosphere \citep{2012MNRAS.426..656S}.
The authors were kind to provide the AD model inference (particularly $L_{5100}$ continuum luminosity) calculated specifically for the observed X-ray luminosity of J1448+16 in low-mass regime ($10^{5}-10^{6}M_{\odot}$), which is shown in Table~\ref{eq: bolometric_cor}. The bolometric luminosity in this case is $L_{bol} = 1.56\times10^{43}$~$\mathrm{erg/s}$. $L_{\rm 2-10\,keV}$ was assumed to be roughly 20\% of $L_{bol}$, due to the requirement that $L_{X-ray}$ is always smaller than $L_{ionization}$; this bolometric correction factor is $\approx$ 3.6 times higher than the previous empirical estimate from \citet{2020A&A...636A..73D}. More accurate estimates of the bolometric luminosity would require the use of the hardness ratio, $\alpha_{OX}$, in future works. 
We display this bolometric luminosity as the purple line (upper one) in Figure~\ref{fig: accretion rate}. 
We calculated the accretion rate constraint as $L_{bol}/L_{Edd} = 104^{+92}_{-50}-112^{+100}_{-54}$~percent using this bolometric luminosity estimate and BH mass estimates from $R_{BLR,~H\alpha}(L_{H\alpha})$~\citep{2023ApJ...953..142C}. The individual uncertainties correspond to individual uncertainties of the BH mass estimation, while the accretion rate range corresponds to different SSP models.

The conceptual problem here is that this AD model~\citep{2012MNRAS.426..656S} is only appropriate for an accretion rate of $L_{bol}/L_{Edd}<0.5$. In general, the classical thin disk model~\citep{1973A&A....24..337S} is often used due to its simplicity, but it is typically invoked for stellar mass BHs. It also has a number of critical failings~\citep{2013LRR....16....1A}. There are a variety of possible replacements for different regimes~\citep[e.g.,][]{2024OJAp....7E..20H}, but how these models relate to AGN SED variations is still widely unknown.
For super-Eddington~\citep{2019BAAS...51c.352B} accretors, the thin disk model should be changed to a slim-disk model \citep{1988ApJ...332..646A} with a disk thickness of $H/R>0.5$. However, the bolometric corrections for the IMBH or low-mass AGN regime, considering the slim disk model, have not been developed yet and will be an aspect of future work.

The accretion rate calculations agree within 68\% confidence interval and point to a relatively high accretion rate ($\gtrsim0.1L_{\rm Edd}$ considering the lower edge of the $L_{bol}$ and higher edge of the $M_{BH}$ confidence intervals), with some possibility of being a super-Eddington accretor. The accretion rates obtained using the H$\alpha$ BH mass estimator, $\mathrm{M_{BH,~R_{C23}}}$, along with both semi-empirical~\citep{2020A&A...636A..73D} and simulated~\citep{2019MNRAS.488.5185N} bolometric conversions, are shown for each SSP model used for full-spectrum fitting in Table~\ref{tab: MBHs_single_epoch_Gaussian}.

\section{Discussion}
\label{sec: discussion}
In the following subsections, we discuss the sources of uncertainties in BLR radius and MBH mass estimations made in this work, the discrepancy between BLR radii derived from RM and R-L relations, and the inspiration for future monitoring campaigns.

\subsection{Sources of uncertainty in BLR radius estimations}

For both the BLR radius and BH mass uncertainty calculation, we used Monte Carlo sampling, incorporating sources of aleatoric uncertainties (uncertainties on the coefficients of linear relations and inner scatters). 
For the light curve analysis, we did not employ the x-axis uncertainty (individual exposure time).
We did not account for the time dilation effect $\sim(1+z)$; for the redshift here, it only leads to a $\sim$ 4\% uncertainty.  
In the following subsections, we discuss the most significant sources of uncertainties and their possible contributions to the BLR radius estimation and, consequently, to the BH mass error budget. 

\subsubsection{Virial factor, $f_{\lambda}$}
In Eq.~\ref{eq: virial_mass}, we specify the virial factor, $f_{\lambda}$, as the factor connecting BH mass, BLR radius, and FWHM for the given wavelength. We have not considered different virial factors for different emission transitions, which can lead to systematic uncertainties of $\sim~10\%$ between H$\alpha$ and H$\beta$ BH mass estimators. 
\cite{2017FrASS...4...70M} and \cite{2020BAAA...61R..87C} found a negative $f_{\lambda}-FWHM^{-1}$ correlation, which can lead to $M_{BH}$ mismatchs by up to a factor of $\sim6,$ compared to alternative BH mass estimates obtained from the properties of the accretion disk emission.
Alternatively, \cite{2024ApJ...963...30L} argued for a positive $f_{\lambda}-L_{bol}/L_{Edd}$ correlation. The $f_{\lambda}$-correction decreases the percentage of high-accreting AGNs in SDSS~DR7 by a factor $\sim100,$ blurring the distinction between high- and low-accreting sources. 
\cite{2024ApJS..274...24Y} demonstrated a 1-2~dex span of $f_{\lambda}$ across a large AGN sample, proposing a connection between the $f_{\lambda}$-correction and different morphological properties of the host galaxy based on spectral parameters.
To summarize, the virial factor remains the main source of uncertainties for MBH in cases where the sphere of BH influence is unresolved.
\subsubsection{Hubble constant and distance uncertainties}
The Hubble constant is determined with an accuracy up to the value of the Hubble tension between local ($H_0=73$~km~s$^{-1}$~Mpc$^{-1}$, \citealt{2022ApJ...934L...7R}) and cosmological ($H_0=67.5$~km~s$^{-1}$~Mpc$^{-1}$, \citealt{2020A&A...641A...6P}) measurements. For this study, we used a mean value $H_0=70$~km~s$^{-1}$~Mpc$^{-1}$. We note that %to stay consistent with \citet{2024ApJ...971..173S}. 
\citet{2013ApJ...767..149B} used $H_{0}=72$~km~s$^{-1}$~Mpc$^{-1}$ from HST Key Project \citep{2001ApJ...553...47F}. We did not normalize the BH mass empirical relations by the Hubble constant. This uncertainty results in a difference between luminosity calculations that can reach $\sim15\%$ at maximum, leading to the $\sim7\%$ uncertainties in the BLR radius and BH mass estimations.

NGC~4051 is a good example of distance accuracy consequences in deriving BH mass and accretion rate, being a very close galaxy that is available for Cepheid distance measurements. \citet{2015ApJ...806...22D} presented a sample of very rapidly accreting BHs with H$\beta$ time lag measurements, where NGC~4051 was claimed to be a super-Eddington IMBH candidate. 
However, \cite{2021ApJ...913....3Y} suggested a 3.2 times higher luminosity due to a previously underestimated distance, along with a lower BLR velocity because it was originally measured as a weighted mean over a few spectra, giving a final estimate that was higher for the BH mass and lower for the accretion rate. This example, however, is more relevant for the galaxies with distances less than 100~Mpc, where peculiar velocities play a significant role. In our case, the distance uncertainties are caused solely by the Hubble tension.

\subsubsection{Choice of calibration stars}
For RM, the sample selection of calibration stars affects the flux calibration and can increase uncertainties. \citet{2011A&A...535A..73H} pointed out that a more accurate flux calibration can be achieved in photometric reverberation mapping campaigns compared to spectral campaigns. This is due to a higher number of calibration stars in use, which unfortunately could enable leaking of variable stars to the sample. It can change the time lag extraction results significantly. In our work, we selected stars by cutting on the SDSS~g\'~magnitude, variability flags from Gaia DR3, S/N, and proper motions criteria (see Section~\ref{sec: flux_calibration} for details).

We opted to use faint validation stars with fluxes similar to or lower than the AGN flux to avoid misconceptions and clearly demonstrate the residual variability remaining after the calibration by brighter stars. 
In Figure~\ref{fig:rel_var_ampl_whole_lc}~(right), we show 54 faint validation stars with small dark cyan crosses, which confirm the high AGN variability amplitude (relative to the flux level) in the narrow H$\alpha$ filter and show a moderate level of AGN variability in the SDSS~g\'~filter probably due to the higher relative contamination of the host galaxy continuum in broadband (see the transmission curves for both filters on top the spectrum in Figure~\ref{fig:spec_nbursts}). 
Additionally, we show the variability metrics for a weighted mean validation star (with median flux uncertainties to keep the same signal-to-noise regime for proper comparison) following the flux calibration (see Table~\ref{tab:var_lc}).

\subsubsection{Choice of a bolometric correction}
% The choice of the bolometric calibrations between X-ray luminosity $L_{X,~2-10keV}$ and bolometric luminosity $L_{bol}$ affects the calculation of the accretion rate.
% We discussed uncertainties in accretion rate calculations in Section~\ref{sec: accretion_rates}.

\begin{table*}[!h]
\centering
\caption{Continuum luminosities at 5100~$\AA$ and BLR radii for the IMBH and low-mass AGN regime. \label{tab: bolometric_corrections}}
\begin{tabular}{lcccccc}
\smallskip
$\log{M_{BH}/M_{\odot}}$ & $L_{bol}/L_{Edd}$ & $L_{5100}$[$\mathrm{erg~s^{-1}}$] & $k_{bol}(5100~\AA)$ & $R_{BLR,~\mathrm{H\alpha}}$(AD face-on)~[$\mathrm{lt-days}$] & $R_{BLR,~\mathrm{H\alpha}}(\theta=45^{\circ})$~[$\mathrm{lt-days}$]\\
\hline
\hline
\smallskip
5.0 & 1.04 & 5.80e40 & 480 & $1.06^{+0.65}_{-0.40}$ & $0.78^{+0.49}_{-0.29}$\\
\smallskip
5.5 & 0.328 & 2.24e41 & 124 & $2.21^{+1.32}_{-0.81}$ & $1.62^{+0.97}_{-0.60}$\\
\smallskip
6.0 & 0.104 & 6.63e41 & 42 & $3.96^{+2.34}_{-1.41}$ & $2.90^{+1.71}_{-1.05}$\\
\hline
\end{tabular}
\tablefoot{
Bolometric corrections $k_{bol}(5100~\AA)$ and $L_{5100}$[$\mathrm{erg~s^{-1}}$] luminosity of the AD  are calculated as an extension of \citet{2019MNRAS.488.5185N} using the \citet{2012MNRAS.426..656S} model. $\theta$ is the inclination of the AD to the line of sight. The AD model adopts a spin parameter = 0.7, and an accretion rate of $2.6\cdot10^{-3}$ $M_{\odot}/\mathrm{year}$, corresponding to $L_{bol} = 1.56\cdot10^{43}~\mathrm{erg/s}$. The BLR radii uncertainties correspond to the 68\% confidence interval and are calculated using Monte Carlo sampling, incorporating the inner scatter and coefficients' uncertainties for the $R_{BLR,~H\beta}(L_{5100})$ relation~\citep{2013ApJ...767..149B}.
}
\end{table*}

The effect of the bolometric calibration choice is connected to the relation between the optical continuum luminosity $L_{5100~\AA}$ and X-ray luminosity $L_{X}$ \citep{2013peag.book.....N}, which we used for the BLR radius expectation to design monitoring in Section~\ref{sec: rc600_campaign}, as well as for the BH mass estimation using the BLR radius derived by the X-ray and H$\beta$ average $R_{BLR} - L_{5100}$ relation \citep{2013ApJ...767..149B} in Section~\ref{subsec: MBHs_by_x_ray}.
We investigated how the particular assumptions (IMBH with high accretion rate) in the bolometric correction calculation can affect the BLR radius predictions from the empirical relation above. To calculate $L_{5100}$ from the observed hard X-ray $L_{X,~2-10~keV}$ for the high-accreting IMBH, we used an extension of the AD model of \citet{2019MNRAS.488.5185N} provided by those authors (see details in Section~\ref{sec: accretion_rates}). Table~\ref{tab: bolometric_corrections} demonstrates simulation results for the face-on AD (highest bolometric luminosity case). For a $\theta=45^{\circ}$ line of sight inclination of the AD, one needs to multiply luminosities by 0.56. We note that the model is uncertain in the case of $L_{bol}/L_{Edd}>0.5$ and for more accurate $L_{bol}$ estimates, it is necessary to know the hardness ratio, $\alpha_{OX}$, and to consider the use of a slim disk model, which will be part of the future work.

%Considering different spectral lines for black hole mass deriving using the single epoch method, one requires to use slightly different R-L relations, because a particular spectral transitions correspond to a particular Stromgren sphere radiuses.

\subsection{Considering whether the BLR size meets our expectations}

%$\citet{2021ApJ...920....9L} indicated that super-Eddington accretor PG~1001+291 does not follow the size–luminosity relation compared to sub-Eddington accretor PG~0923+201.
From Table~\ref{tab: bolometric_corrections}, we notice the importance of using appropriate assumptions for the AD modeling to predict the BLR radius, especially for high accretion rate and IMBH regimes. The initial assumptions for the BLR RM cadence estimation (see the monitoring design in Section~\ref{sec: design_blr_rm}) considered this AGN as a moderate accretion rate source and predicted a possible longer outcome for the time lag. 
With the use of ICCF, we derived a tentative time lag of $1-8$ days in section~\ref{subsubsec: CCF}. The lower limit of time lag $\tau_{low} = 1~\mathrm{days}$ is similar to the predicted BLR radiuses for $\log M_{BH}/M_{\odot} = 5.0$ from Table~\ref{tab: bolometric_corrections}; however, within the error bars, it could match all BLR radii predictions listed in Table~\ref{tab: bolometric_corrections}. 

The overall cadence of the present RM campaign does not efficiently sample very short BLR radii smaller than 1~ld. However, we could investigate the intra-night timescale of variability on the night of 2024-06-13, which is significant in both filters ($\sim$1.7~hours mentioned in Section~\ref{sec: intra-night}).  We also note a significant correlation in the temporal flux variability pattern in both filters during $\sim$~1~hour. Such a fast variability can be seen as an additional signature of the IMBH regime.

We inferred two possible natures of such a short timescale variability in the narrow H$\alpha$ filter, which are summarized in Table~\ref{tab: scenarios}: 
(a) the source of intra-night variability in the narrow H$\alpha$ band is BLR emission, thereby making the AD and NLR contributions  negligible. Then $\sim$1.7~hours would correspond to the AD-to-inner-BLR light travel time (the same pattern of intra-night variability in SDSS~g\'~and narrow H$\alpha$ is a coincidence). However, our RM campaign did not test for such short BLR-radii due to insufficient cadence; 
(b) alternatively, the $\sim$1.7~hour timescale of variability could originate directly from AD continuum contribution to the narrow H$\alpha$ filter (as discussed in Section~\ref{sec: design_blr_rm}). Then the auto-correlation term would appear in the CCF,  changing the centroid of the recovered AD-to-BLR delay~\citep{2012A&A...545A..84P}. %The (a) case is less likely to fit the approximately same timescale of intra-night variability in both SDSS~g\'~and narrow H$\alpha$ filters. 
%The (b) case can fit such a setup.
 %This hints to that the AD emission variability is seen directly in the H$\alpha$ filter. 
%Then the BLR should be `transparent' to show `naked' AD emission.
Then the BLR would be "transparent" (optically thin along the line of sight) on such short timescales of < 1 day, allowing for the accretion disk emission to be directly visible. %However, the strong relative variability of this filter flux is probably also due to variability in the broad and narrow emission lines at the same time. 
Such behavior is plausible, for instance in the case of a $\sim$~1~hour recombination time ($\tau_{rec}\sim40\cdot(10^{11}~\mathrm{cm}^{-3}/n_{e})~\mathrm{s}$;~\citealt{2006LNP...693...77P}) along the line-of-sight part of the BLR due to the relatively "low" average electron density ($n_{e}\sim10^{9}~\mathrm{cm}^{-3}$;~\citealt{2006agna.book.....O}). We notice that on the physical and temporal scales of BLRs expected in IMBHs (tens of minutes to days), we could reach a breaking point for key assumptions of BLR RM, such as the ``nearly instantaneous'' (minutes-to-hours compared to days-to-weeks) ionization and recombination time~\citep{1993PASP..105..247P,2006LNP...693...77P}. 

The ``transparency''  of the BLR on short timescales could be explained by the geometry, %Assuming a spherical BLR structure, we can draw a rough parallel with emission within HII regions (though the electron densities in BLRs are much higher~\citealp{2006agna.book.....O}), 
where the central part in the observer direction is optically thin. 
For instance, in self-shielding BLR models, the column density increases toward the equatorial plane, creating a structure where ionizing radiation escapes more easily in the polar direction. This results in a vertically extended, stratified distribution resembling a ``bird's nest'' when viewed face-on, with line emission arising predominantly from the surface layers of the flattened distribution~\citep{2007arXiv0711.1025G}.
%While the column density increases towards the edges, the whole structure likely emits as a shell, which can be, in the case, e.g., the self-shielding model -- face-on 'bird's nest' geometry of BLR structure~\citep{2007arXiv0711.1025G}. %reformulate to avoid amplitude discussion
 
To explain higher amplitude variability in narrow H$\alpha$ compared to SDSS~g\'~in case (b), we consider filter widths. Considering the thermal reprocessing model, we observe the change in outer AD temperature with radius as a simple blackbody variation ~\citep{1969Natur.223..690L,1973A&A....24..337S}. Then, as a first approximation, different parts of outer AD emit at different wavelengths $R\sim\lambda^{4/3}$ according to a thin-disk temperature profile $T(R)\sim R^{-3/4}$~\citep{1973A&A....24..337S} and Wien's law. %$T\sim\frac{b}{\lambda}$.
Therefore, we point out that SDSS~g\'~ is a broad filter and we are averaging the signal over a spatially larger part of the AD than for the AD area in the narrow H$\alpha$ wavelength range. The ratio of spatial AD areas, which emit in broad SDSS~g\'~and narrow H$\alpha$ filters correspondingly:  
\begin{equation}
    \frac{\Delta R_{SDSS~g~}}{\Delta R_{H\alpha}} \sim  (\frac{\lambda_{SDSS~g}}{\lambda_{H\alpha}})^{1/3} \frac{\Delta\lambda_{SDSS~g}}{\Delta\lambda_{H\alpha}}\approx 13,
\end{equation}
where $\Delta\lambda_{SDSS~g}$ and $\Delta\lambda_{H\alpha}$ are the filter widths, while $\lambda_{SDSS~g}$ and $\lambda_{H\alpha}$ are the central wavelengths of the corresponding filters. This wavelength-averaging effect could damp down the amplitude of variability in SDSS~g\'~\citep{2005MNRAS.360..816Z} at a factor of $\sim\frac{1}{\sqrt{\text{number~of~zones}}}\approx3.5$ based on the assumption of incoherent-phase variability, which is the same in each spatial zone of AD, where the flux of each considered zone is roughly the same. 
Additional possibilities to amplify variability in narrow H$\alpha$ are 1) case (a) friendly:  simple overestimation of AD contribution to the whole flux in SDSS~g\'~filter (< than the contribution of BLR to narrow H$\alpha$), along with a gravitational microlensing amplification of the BLR~\citep{1988ApJ...335..593N} or/and BLR breathing, with the responsivity increasing towards bigger BLR radii~\citep{2014MNRAS.444...43G}; 2) case (b) friendly: the variable BLR and NLR components on top of the AD variability.   

From a gravitational perspective, BLR clouds can be considered as either gravitationally supported or (more likely) pressure-supported objects~\citep[see, e.g., chapter 5.6 in][]{2013peag.book.....N}. For the rough estimation of the minimal orbit of BH-bounded BLR clouds in a self-gravitationally supported case, we pick the Roche limit. For an individual BLR cloud, the Roche limit is $\sim 0.58~[\text{lt-days}],$ assuming $M_{BH}\sim10^{5}M_{\odot}$ and given the critical electron density of $n_{c}\sim8.7\times10^{16}\mathrm{cm^{-3}}$~\citep{2006agna.book.....O} for Ly$\alpha$ photons, which must escape for the ionization of the next layers of hydrogen in the case B recombination process. Therefore, this hints at the possibility that the intra-night variability in narrow H$\alpha$ does not originate from a compact BLR; this would refer to case (a) if BLR clouds are internally supported by self-gravitation.
In a pressure-supported case, the outward pressure force of a BLR cloud is significantly stronger than the force of self-gravitation. The BLR clouds are confined externally by the magnetic field pressure, radiative pressure, or winds~\citep{2013peag.book.....N}. We are currently not able to speculate about the exact BLR confinement mechanisms in J1448+16, leaving this matter to a future work. 

From a photoionization perspective, \citet{2025ApJ...980..134W} used {\sc Cloudy} to check how the BLR radii (time delays) change position on the conventional $R_{BLR}-L^{0.5}$ relation in the case of a relatively high accretion rate. They indicated that {\sc Cloudy} predictions for time delays are larger in the case of high accretors, which does not follow the observable trend of smaller delays for such objects~\citep{2023FrASS..1030103P,2024ApJS..275...13W}. Such a discrepancy could be explained by the self-shielding winds of thicker ADs than the classical thin AD model of \citet{1973A&A....24..337S}, or/and the electron density gradient inside of the BLR. 
We also propose a scenario in which the ionization radius is bigger than the virialized BLR radius: $R_{ionized} > R_{BLR}$. In such a case, we would observe variability in both the BLR and the NLR during monitoring. Then, both the high amplitude of variability in narrow H$\alpha$ filter and spread in the CCF could be explained by the contribution of variable NLR and the corresponding additional auto-correlation component, which reduces the derived time lag in case of insufficient separation of auto-correlation and correlation peaks~\citep{2012A&A...545A..84P}. The quantifying tests of this effect will be considered as part of a future study.

The current observations and simulations are very uncertain for low-mass AGNs and allow only for loose constraints to be set on the BLR size, BH mass, and accretion rate. This fact strengthens our motivation to calibrate the R-L relations at the faint end. High cadence RM campaigns for J1448+16 and its siblings are essential for understanding active IMBH properties.
\subsection{Prospects for future IMBH RM campaigns}

The dust RM for highly accreting IMBHs can constrain the dust radius in this extreme condition of high radiative pressure during much shorter monitoring (a couple of months) than SMBHs (a couple of years to decades). 

The AD RM for IMBH is a fast (a couple of nights) monitoring technique, which can test the AD transfer function and provide insights into the dominant mechanism of AD emission in highly accreting AGN cases.
We took advantage of ``time-squeezing'' machines and started dust and AD RM campaigns for J1448+16 and other rapidly accreting IMBH candidates.

We notice the potential caveats for future IMBH RM campaigns, which are contrary to the conventional RM assumptions: (a) instantaneous ionization and recombination time of BLR clouds~\citep{2006LNP...693...77P} could be breaking, because the light travel time would not be the most important timescale; %(b) smaller NLR size can affect long campaigns by the additional delay in the narrow H$\alpha$ passband signal; 
(b) the BLR structure is stable throughout RM campaign~\citep{2006LNP...693...77P} could be breaking, because the high-accreting rate IMBHs are on small BLR dynamical scales (weeks-to-years).  
\section{Conclusions}
\label{sec: conclusion}
We present narrow H$\alpha$ and broad SDSS~g\'~photometric monitoring and XMM-Newton spectroscopy of the IMBH candidate J1448+16. We confirm J1448+16 as an active IMBH based on a significant point-source X-ray detection. The 2-10~keV X-ray luminosity is $L_{X}[2-10~\mathrm{keV}] = \mathrm{3.33^{+0.49}_{-0.44}\times10^{41}erg~s^{-1}}$, with a power-law photon index of~$\Gamma = 2.32^{+0.15}_{-0.13}$ hinting at a high accretion rate. Using the FWHM derived from SDSS full-spectrum fitting, we estimated the single-epoch BH mass of J1448+16 to be $M_{BH} = 0.9^{+0.9}_{-0.5}-2.4^{+3.7}_{-1.4}~\times10^{5} M_{\odot}$, using broad H$\alpha$ luminosity (or $M_{BH} = 1.0^{+1.2}_{-0.5}-1.3^{+0.7}_{-0.5}~\times10^{5} M_{\odot}$ using X-ray luminosity). Both estimates are in agreement, placing this object in the IMBH regime.

We confirm J1448+16 as a rapidly accreting IMBH by measuring the accretion rate using average-AGN semi-empirical~\citep{2020A&A...636A..73D} and simulated~\citep{2019MNRAS.488.5185N} bolometric corrections.
Both bolometric luminosities ($L_{bol} = 5.56\times10^{42} \mathrm{erg/s}$ and $L_{bol} = 1.56\times10^{43} \mathrm{erg/s}$, respectively) place J1448+16 well within the rapidly accreting regime, with the possibility of it being a super-Eddington accretor ($L_{bol}/L_{Edd} = 37^{+66}_{-24}-40^{+72}_{-26}$~per~cent and $L_{bol}/L_{Edd} = 104^{+92}_{-50}-112^{+100}_{-54}$~per~cent). 

We performed a proof-of-concept intra-night variability study for a high-accretion-rate IMBH (previous studies did not detect AD variability for this type of source~\citep{2024ApJ...974..288Z}). We demonstrated the feasibility of BLR RM for rapidly accreting IMBH identification and verification, which is especially significant in the ELT era.

In addition, we extracted a tentative time lag of 1-8 days, which might be caused by the sampling window. 
We notice a possible discrepancy between the BLR radius derived from an upper limit of $\tau_{large}\sim8~\mathrm{days}$ estimation and $R_{H\beta}-L_{5100}$~\citep{2013ApJ...767..149B} with $R_{H\alpha}-R_{H\beta}$~\citep{2023ApJ...953..142C}; along with $R_{H\alpha}-L_{H\alpha}$~\citep{2023ApJ...953..142C} relations. The larger BLR radius than expected from the extrapolation of empirical relations indicates the possibility that this case deviates from the classical thin disk model due to self-shielding winds of thicker AD or electron density gradient in the BLR \citep{2025ApJ...980..134W}.

For the first time, we were able to detect a significant (as high as $\pm55\%$ in H$\alpha$) intra-night variability on a timescale of $\sim 1.7$~hours in a rapidly accreting IMBH using the 60cm telescope RC600, which is strengthened by the correlation of the SDSS g\'~and narrow H$\alpha$ fluxes. This finding demonstrates the remarkable capability of small-aperture instruments in IMBH studies. 
The detected high-amplitude intra-night variability and extracted time lag in rapidly-accreting IMBH are promising for future validation studies using 6-8~m telescopes with much smaller individual exposure times and a higher cadence.

\bibliographystyle{aa}
\bibliography{IMBHs}

\begin{appendix}

\onecolumn
\section{}
\begin{acknowledgements}
MD expresses gratitude to Hagai Netzer (for simulation of particular bolometric corrections as extension of the \citep{2019MNRAS.488.5185N}), Francisco Pozo (for active discussions of time lag estimation), Kai Polsterer (for structure discussion), Nikos Gianniotis (for GPCC discussion), Ren\'{e} Andrae (for discussion of Monte Carlo sampling), Nadine Neumayer (for structure discussion), Christian Fendt (for AD models and Roche limit discussion), Nico Winkel (for profiles of broad lines discussion), Jochen Heidt (for being thesis committee member), Irina Smirnova-Pinchukova (for super-Eddington IMBHs discussion), Dhruv Muley (as native-speaker editor).
We thank the anonymous referee for comments and suggestions, which improved the clarity of the work presentation.
The RC600 CMO telescope was created as part of the M.V. Lomonosov Moscow State University Program of Development. 
Part of this work was done during the MD's internship at the Optical and Infrared Astronomy division of the Center for Astrophysics --- Harvard and Smithsonian.
MD also acknowledges the Hubble Space Telescope grant HST-GO-16739 (PI: IC), which covered the internship in the Optical and Infrared (OIR) Division of the Smithsonian Astrophysical Observatory.
AA was supported by a European Research Council (ERC) grant UniverScale, agreement 95154.
ER was supported by the Russian Science Foundation (RSCF) grant No.~23-12-00146. IC's research is supported by the Telescope Data Center at the Smithsonian Astrophysical Observatory. IC also acknowledges the support from the NASA ADAP-22-0102 grant (award 80NSSC23K0493) and NASA XMM-Newton Data Analysis grant (award 80NSSC22K0389). 
FEB acknowledges support from ANID-Chile BASAL CATA FB210003, FONDECYT Regular 1241005, and Millennium Science Initiative AIM23-0001.
This research has made use of the SIMBAD database, operated at CDS, Strasbourg, France.
This work made use of Astropy, a community-developed core Python package and an ecosystem of tools and resources for astronomy \citep{astropy:2013, astropy:2018, astropy:2022}, particularly Photutils \citep{larry_bradley_2023_7946442}, and in addition Python packages SciPy \citep{2020SciPy-NMeth} and {\tt\string scikit-learn} \citep{scikit-learn}.
This research used TOPCAT, a software tool for astronomical data analysis \citep{Taylor2005} and Flexible Image Transport System (FITS) data format \citep{2001A&A...376..359H}. We complement the use of LLM-based tools (ChatGPT for finding appropriate wording and references in some cases, Grammarly for grammar and spelling convention checks) with critical thinking, as was proposed in \citep{2024arXiv240920252F}. 
Based on observations obtained with XMM-Newton, an ESA science mission with instruments and contributions directly funded by ESA Member States and NASA.
This paper is dedicated to the memory of Pasha Tekhnik and David Lynch. 
\end{acknowledgements}

\section{Location of reference stars on the FoV}
\begin{figure}[b!]
\centering
\includegraphics[width=0.81\linewidth]{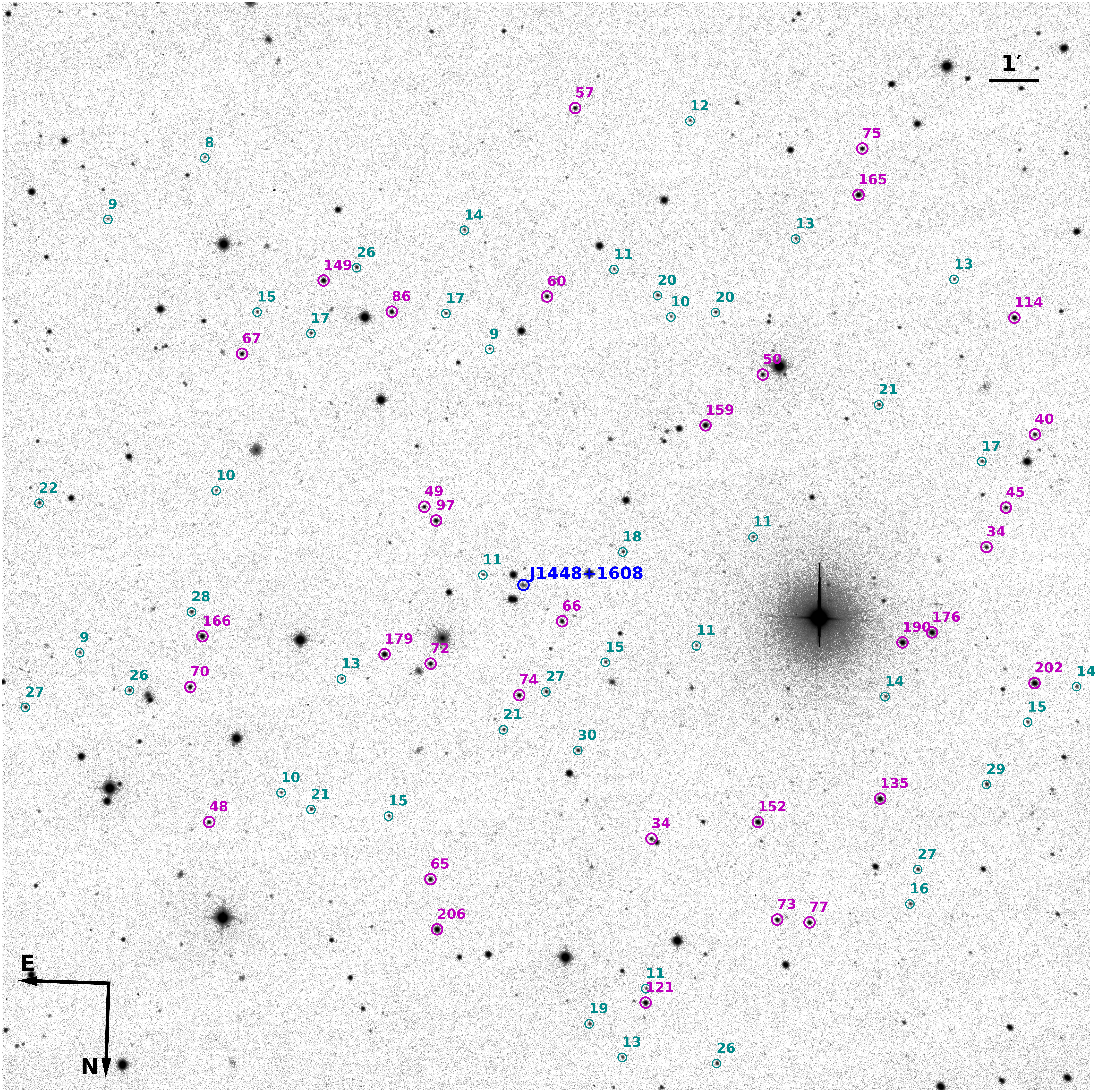}

\caption{Example of an RC600~g\'~field (FoV 22.7~arcmin) with reference stars marked with magenta circles and validation stars with dark cyan circles, indicating a S/N near each star. The circle diameter is 6'' for display purposes (the photometry aperture diameter is 3''). 
\label{fig:foV}} 
\end{figure}

\twocolumn
\newpage
\clearpage
\section{Description of AGN light curve columns}
In this appendix, we describe the metadata of the supplementary Table~\ref{tab: light_curve_agn}, made available in electronic form at the CDS portal via anonymous ftp to cdsarc.cds.unistra.fr (130.79.128.5) or via \url{https://cdsarc.cds.unistra.fr/cgi-bin/qcat?J/A+A/}.

\begin{table}
\centering
\caption{Description of AGN light curve columns of a supplementary table.
\label{tab: light_curve_agn}}
\begin{tabular}{lccc}
\hline
\hline
Column Name & Units & Datatype & Description \\
\hline
\hline
Band & & string & Photometric band (g or H$\alpha$) \\
Photon count & ADU & float & Number of extracted photons in 3'' aperture before flux calibration\\
Flux & $10^{-17}$ erg s$^{-1}$ cm$^{-2}$ & float & AGN flux\\
Flux\_err & $10^{-17}$ erg s$^{-1}$ cm$^{-2}$ & float & Error of the AGN flux\\
MJD & days & float & Modified Julian Date\\
Correction & $10^{-17}$ erg s$^{-1}$ cm$^{-2}$ & float & Relative correction term (median variation of the calibration star)\\
Airmass &  & float & Airmass during the individual exposure\\
Humidity & percent & float & Humidity during the individual exposure\\
\hline
\hline
\end{tabular}
\end{table}

\section{Intra-night variability scenarios}
\begin{table}[!h]
\centering
\caption{Possible scenarios for intra-night variability and how they fit observational constraints.} \label{tab: scenarios}
\begin{tabular}{p{5.55cm}|p{4cm}|p{3cm}|p{3cm}}
\backslashbox{Possible scenario}{Observational fact} & Variability amplitude in narrow H$\alpha$~>~SDSS~g\'~ & Similar temporal pattern & Timescale $\sim1.7$~h in H$\alpha$\\
\hline
\hline
\smallskip
 Case (a): dominant source of narrow H$\alpha$ variability is BLR; AD and NLR contributions are negligible.  & \begin{enumerate}
    \itemsep0em
     \item Contribution of AD to SDSS~g\'~ is < contribution of BLR to narrow H$\alpha$ ($\sim$20\%).
     \item BLR can be amplified by gravitational microlensing~\citep{1988ApJ...335..593N}.
     \item BLR is breathing; responsivity is increasing towards bigger BLR radii~\citep{2014MNRAS.444...43G}.
 \end{enumerate}  & \smallskip Coincidence of the same phase & \smallskip The variability timescale $\sim$~outer AD size;  The minimal possible AD-to-inner-BLR light travel time as a consequence of the same phase (can be N timescales bigger).\\
 \hline
\smallskip
 Case (b): dominant source of narrow H$\alpha$ variability is AD; BLR and NLR also contribute. & \begin{enumerate}
     \itemsep0em
     \item The difference in widths of filters (integration over wavelength means averaging amplitude over different spatial AD regions $\Delta R_{SDSS~g} \sim 13 \times \Delta R_{H\alpha}$)~\citep{2005MNRAS.360..816Z}.
     \item Optically thin BLR on the line-of-sight, the addition of BLR and NLR variability trends on top of AD variability.
 \end{enumerate}  & \smallskip AD is dominant in both filters; delays between different parts of AD are smaller than the individual exposure time.  & \smallskip Light travel time between different parts of outer AD.\\
\hline
\end{tabular}
\end{table}
\newpage
\clearpage
\section{Black hole mass estimation using single-epoch virial calibrations}
\label{sec: appendix_A_single_epoch_mbhs}
In this appendix, we describe single-epoch BH mass estimators based on observed broad H$\alpha$ emission line, particularly, its luminosity and FWHM.
Table~\ref{tab: M_BH_spectra} shows stellar population parameters, emission line fluxes, and FWHM derived using full-spectrum fitting with an assumed Gaussian BLR profile (discussed in Section~\ref{subsec:sdss_spec}).

The ``conventional'' (484 citations up to date according to adsabs) BH mass estimator of \citet{2013ApJ...775..116R} combines three empirical relations: $R_{BLR,~H\beta}(L_{5100})$ from \citet{2013ApJ...767..149B}, $L_{H\alpha}(L_{5100})$ and $\text{FWHM}_{H\beta}(\text{FWHM}_{H\alpha})$ from \citet{2005ApJ...630..122G} for virial BH mass estimator through broad H$\beta$ emission line:
\begin{equation}
    M_{BH,~H\beta} \equiv M_{BH,~R_{B13}} = f_{H\beta}\cdot\frac{R_{BLR,~H\beta}\cdot\text{FWHM}_{H\beta}^{2}}{G},
\label{eq: mbh_hbeta}
\end{equation}
where the virial factor $f$:
\begin{equation}
f_{H\beta}=1,
\end{equation}
and where $L_{H\alpha}$ is connected with $L_{5100}$ as:
\begin{equation}
    L_{\mathrm{H}\alpha} = (5.25 \pm 0.02) \times 10^{42} \left( \frac{L_{5100}}{10^{44} \ \mathrm{erg \ s}^{-1}} \right)^{(1.157 \pm 0.005)} \ \mathrm{erg \ s}^{-1},
\end{equation}
with an intrinsic scatter of $\sigma_{int} = 0.2$~dex.
$\text{FWHM}_{\text{H}\beta}$ is related to the $\text{FWHM}_{\text{H}\alpha}$ as
\begin{equation}
\text{FWHM}_{\text{H}\beta} = (1.07 \pm 0.07) \times 10^3 \left( \frac{\text{FWHM}_{\text{H}\alpha}}{10^3 \, \text{km s}^{-1}} \right)^{(1.03 \pm 0.03)} \, \text{km s}^{-1},
\label{eq: fwhms}
\end{equation}
with an intrinsic scatter of $\sigma_{int}=0.1$~dex.

$R_{BLR,~H\beta}$ was calculated directly by using $L_{5100}$ as an input of the relation from \cite{2013ApJ...767..149B} given by
\begin{equation}
\log\left(\frac{R_{BLR,~H\beta}}{1~\mathrm{lt-day}}\right)=1.555^{+0.024}_{-0.024}+0.542^{+0.027}_{-0.026}\cdot\log_{10}\left(\frac{\lambda L_{\lambda}}{10^{44}\mathrm{erg~s}^{-1}}\right),
\label{eq: hbeta_r_blr_l5100}
\end{equation} with an intrinsic scatter of $\sigma_{int}=0.19\pm0.02$~dex.

$R_{BLR,~H\alpha}$ can be linked to $R_{BLR,~H\beta}$ following \citet{2023ApJ...953..142C} as $R_{BLR,~H\alpha} = 1.68\cdot R_{BLR,~H\beta}$ with intrinsic scatter $\sigma_{int}= 0.23 \pm 0.03$~dex.

FWHM is the same for all broad components of emission lines in {\sc NBursts} full-spectrum fitting. 
We built the H$\beta$ BH mass estimator, using virial equation (Eq.~\ref{eq: virial_mass}) and $R_{BLR,~H\beta}(L_{5100})$ (Eq.~\ref{eq: hbeta_r_blr_l5100}), avoiding $\mathrm{FWHM_{H\beta}(FWHM_{H\alpha})}$ relation (Eq.~\ref{eq: fwhms}).

However, \citet{2023ApJ...953..142C} claim that this approach to BH mass estimation systematically over-predicts BH masses in the IMBH and low-mass AGN regime by up to 0.7~dex, because of the use of a non-unity slope in $L_{5100}(L_{H\alpha})$ relation~\citep{2005ApJ...630..122G}.

\citet{2023ApJ...953..142C} do not use external $L_{5100}(L\alpha)$ luminosity and FWHM conversions because they measure $R_{BLR,~H\alpha}$ using RM and directly build $R_{BLR,~H\alpha}(L_{H\alpha})$ relation.
\citet{2023ApJ...953..142C} give the $H\alpha$ BLR radius, $R_{H\alpha}$, by the following equation,
\begin{equation}
    \log_{10}\left(\frac{R_{H\alpha}}{1[\mathrm{lt-day}]}\right)=\left(1.16\pm0.05\right)+(0.61\pm0.04)\cdot\log_{10}\left(\frac{L_{H\alpha}}{10^{42}[\frac{\mathrm{erg}}{\mathrm{s}}]}\right)
    \label{eq:cho_2023_t_blr}
\end{equation}
with intrinsic scatter $\sigma_{int} = 0.28\pm0.03$.
We construct a H$\alpha$ BH mass estimator precise up to the virial factor ($f = 1$), based on the virial equation for the BH mass estimation (Eq.~\ref{eq: virial_mass}) and radius of BLR (Eq.~\ref{eq:cho_2023_t_blr}), expressed as
\begin{equation}
M_{BH,~H\alpha} \equiv M_{BH,~R_{C23}} = f_{H\alpha}\cdot\frac{R_{BLR,~H\alpha}\cdot\text{FWHM}_{H\alpha}^{2}}{G}.
\label{eq: mbh_ha_virial}
\end{equation}

For the uncertainty propagation, we used Monte Carlo sampling, incorporating all reported coefficient uncertainties and inner scatters of empirical relations.
We calculate BH masses as the median of the Monte Carlo sampled distribution and uncertainties within 68\% confidence interval using both H$\beta$ and H$\alpha$ BLR radii and demonstrate them in Table~\ref{tab: MBHs_single_epoch_Gaussian}.
The accretion rate estimate using empirical bolometric conversion~\citep{2020A&A...636A..73D} and $M_{BH,~R_{C23}}$ is shown in Table~\ref{tab: MBHs_single_epoch_Gaussian} in the $\epsilon_{D20}$ column. The accretion rate estimate using the extension of \citet{2019MNRAS.488.5185N} and $M_{BH,~R_{C23}}$ is demonstrated in Table~\ref{tab: MBHs_single_epoch_Gaussian} in the $\epsilon_{N19}$ column. The accretion rate uncertainties within 68\% confidence interval are calculated using Monte Carlo error propagation, considering the BH masses and bolometric luminosities distributions. Within uncertainties, accretion rates agree and point to a rapidly accreting regime of IMBH with the possibility of a super-Eddington regime.
\begin{figure}
\centering
\includegraphics[width=0.75\linewidth]{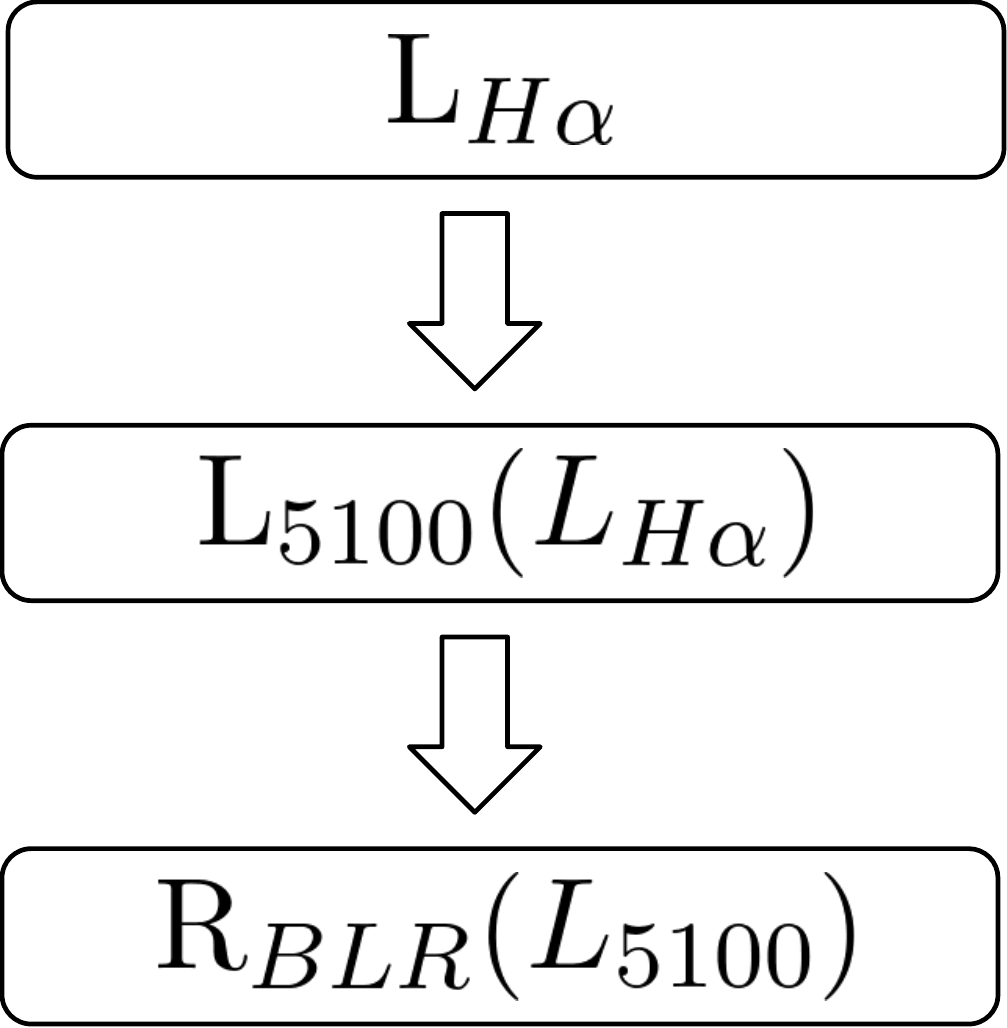}
\caption{General pipeline of deriving $\mathrm{H\beta}$~\citep{2013ApJ...767..149B} BLR radius in this appendix. \label{fig: r_blr_from_xray}}
\end{figure}

\begin{figure}
\centering
\includegraphics[width=0.75\linewidth]{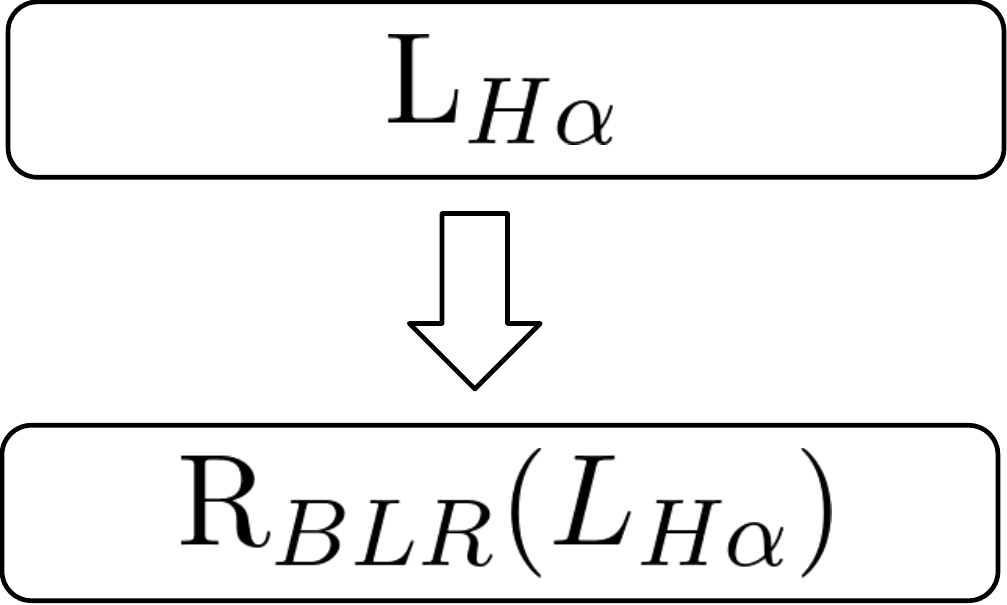}
\caption{General pipeline of deriving $\mathrm{H\alpha}$ BLR radius~\citep{2023ApJ...953..142C} in this appendix. \label{fig: r_blr_from_xray}}
\end{figure}

\begin{landscape}
\begin{table}
\centering
\begin{threeparttable}
\centering
\caption{Parameters for BH mass estimations from single-epoch optical spectrum. 
}
\label{tab: M_BH_spectra}
\begin{tabular}{lcccccccc}
\smallskip
Model & $ \sigma_{*}$, km/s & Age$_{*}$, Gyr & [Fe/H], dex & $ \sigma_{NLR}$, km/s & $ Flux_{\mathrm{[OIII]}}, \frac{erg}{s \cdot cm^2}$ & $ \sigma_{BLR}$, km/s & $ Flux_{\mathrm{H_\alpha}}, \frac{erg}{s \cdot cm^2}$ & $ \chi^2 $ \\
\hline
\hline
E-MILES &  $ 20.71 \pm 1.04 $  &  $ 13.37 \pm 1.12 $  &  $ -0.244 \pm 0.072 $  &  $ 53.0 \pm 11.1 $  &  $ 273.5 \pm 4.8 $  &  $ 299.4 \pm 10.4 $  &  $ 401.0 \pm 11.9 $  & 1.09 \\
MILES (+MgFe) &  $ 20.71 \pm 1.04 $  &  $ 10.49 \pm 1.45 $  &  $ -0.013 \pm 0.083 $  &  $ 53.5 \pm 10.4 $  &  $ 271.4 \pm 4.7 $  &  $ 301.4 \pm 10.4 $  &  $ 393.8 \pm 11.7 $  & 1.08 \\
MILES &  $ 20.71 \pm 1.04 $  &  $ 6.18 \pm 1.48 $  &  $ -0.142 \pm 0.130 $  &  $ 53.4 \pm 10.8 $  &  $ 271.1 \pm 4.7 $  &  $ 305.0 \pm 10.6 $  &  $ 395.3 \pm 11.7 $  & 1.08 \\
PEGASE &  $ 29.3 \pm 12.4 $  &  $ 13.10 \pm 2.37 $  &  $ -0.221 \pm 0.107 $  &  $ 52.8 \pm 10.6 $  &  $ 256.2 \pm 4.4 $  &  $ 299.4 \pm 10.5 $  &  $ 369.5 \pm 11.1 $  & 1.07 \\
X-Shooter &  $ 67.3 \pm 8.0 $  &  $ 14.28 \pm 2.26 $  &  $ -0.142 \pm 0.097 $  &  $ 53.2 \pm 10.3 $  &  $ 275.0 \pm 4.8 $  &  $ 293.3 \pm 10.3 $  &  $ 399.3 \pm 12.1 $  & 1.12 \\ 

\hline
\hline
\end{tabular}
\tablefoot{$Flux_{\mathrm{H\alpha}}$ represents flux of H$\alpha$ broad component. We assume a Gaussian BLR profile with the same FWHM for all detected broad lines.}
\label{tab:spec_nbursts}
\end{threeparttable}
\end{table}

\begin{table}
\centering
\caption{Black hole mass and BLR radii $R_{BLR,~H\alpha}$, $R_{BLR,~H\beta}$ estimations using single epoch virial calibrations.  \label{tab: MBHs_single_epoch_Gaussian}}
\begin{tabular}{lcccccccc}
\smallskip
SSP Model & $\text{FWHM}_{H\alpha}$,~km/s & $L_{H\alpha}, \times10^{40}~\frac{erg}{s}$ & $R_{H\beta,~B13,~L_{H\alpha}}[\mathrm{lt-days}]$ & $M_{BH,~R_{B13}}, \times10^{6}M_{\odot}$ & $R_{H\alpha,~C23,~L_{H\alpha}}[\mathrm{lt-days}]$ & $M_{BH,~R_{C23}}, \times10^{6}M_{\odot}$ & $\epsilon_{D20}$ & $\epsilon_{N19}$ \\ 
\hline
\hline
\smallskip
E-MILES & $ 705.03 \pm 24.49 $ & $ 1.27 \pm 0.04 $ & $ 2.14_{-1.07}^{+2.12} $ & $0.23_{-0.14}^{+0.36}$ & $1.01_{- 0.49}^{+0.96}$ & $0.10_{-0.05}^{+0.09}$ & $0.38^{+0.68}_{-0.25}$ & $1.07_{-0.51}^{+0.94}$ \\
\smallskip
MILES(+MgFe) & $ 709.74 \pm 24.49 $ & $ 1.25 \pm 0.04  $ & $ 2.12_{-1.06}^{+2.1} $ & $0.23_{-0.14}^{+0.36}$ & $1.00_{-0.48}^{+0.95}$ & $0.10_{-0.05}^{+0.09}$ & $0.38^{+0.68}_{-0.24}$ & $1.07 _{- 0.51 } ^{+ 0.94 }$ \\
\smallskip
MILES & $ 718.22 \pm 24.96 $ & $ 1.25 \pm 0.04 $ & $ 2.13_{-1.06}^{+ 2.11} $ & $0.24_{-0.14}^{+0.37}$ & $1.00_{- 0.48}^{+0.95}$ & $0.10_{-0.05}^{+0.09}$ & $0.37^{+0.66}_{-0.24}$ & $1.04 _{- 0.50} ^{+ 0.92 }$\\
\smallskip
PEGASE & $ 705.03 \pm 24.73 $ & $ 1.17 \pm 0.03 $ & $ 2.06_{-1.03}^{+2.04} $ & $0.22_{-0.13}^{+0.35}$ & $0.96_{-0.46}^{+0.92}$ & $0.09_{-0.04}^{+0.09}$ & $0.40^{+0.72}_{-0.26}$ & $1.12 _{- 0.54 } ^{+ 1.00 }$\\
\smallskip
X-Shooter & $ 690.67 \pm 24.25 $ &  $ 1.27 \pm 0.04 $ & $ 2.14_{-1.0}^{+2.12} $ & $0.22_{-0.13}^{+0.34}$ & $1.0_{-0.48}^{+0.96}$ & $0.09_{-0.05}^{+0.09}$ & $0.40^{+0.71}_{-0.26}$ & $1.12 _{- 0.53 } ^{+ 0.99 }$\\ 
\hline
\end{tabular}
\tablefoot{
$M_{BH,~H\beta}(\text{FWHM}_{H\alpha},~L_{H\alpha})$ is defined as $M_{BH,~R_{B13}}$~\citep{2013ApJ...767..149B} and $M_{BH,~H\alpha}(\text{FWHM}_{H\alpha},~L_{H\alpha})$ id defined as $M_{BH,~R_{C23}}$~\citep{2023ApJ...953..142C}. Accretion rate estimates $\epsilon_{D20}$ and $\epsilon_{N19}$ correspond to $L_{bol}/L_{Edd}$, where Eddington luminosity $L_{Edd}$ calculated assuming $M_{BH,~R_{C23}}$~\citep{2023ApJ...953..142C} H$\alpha$ BH mass estimate to use the smallest number of empirical relations altogether, while $L_{bol}$ calculated using bolometric conversions from $L_{X}[2-10keV]$ from \citet{2020A&A...636A..73D} and \citet{2019MNRAS.488.5185N} respectively. Monte Carlo sampling was used for error propagation, incorporating all reported in relation coefficients' errors and inner scatters. BH mass estimators are calculated as the median, and BH mass error bars are calculated as 68\% confidence interval. Luminosity is calculated assuming $z=0.038$, $H_{0}=70$~km~s$^{-1}$~Mpc$^{-1}$. The Gaussian profile is assumed for broad emission lines in the full-spectrum fit.
}
\end{table}

\end{landscape}

\section{Black hole mass estimation using X-ray[2-10keV] luminosity, bolometric correction $L_{5100}(L_{X})$, and virial calibration}
\label{sec: appendix_B_xray_mbhs}

\begin{figure}
\centering
\includegraphics[width=0.75\linewidth]{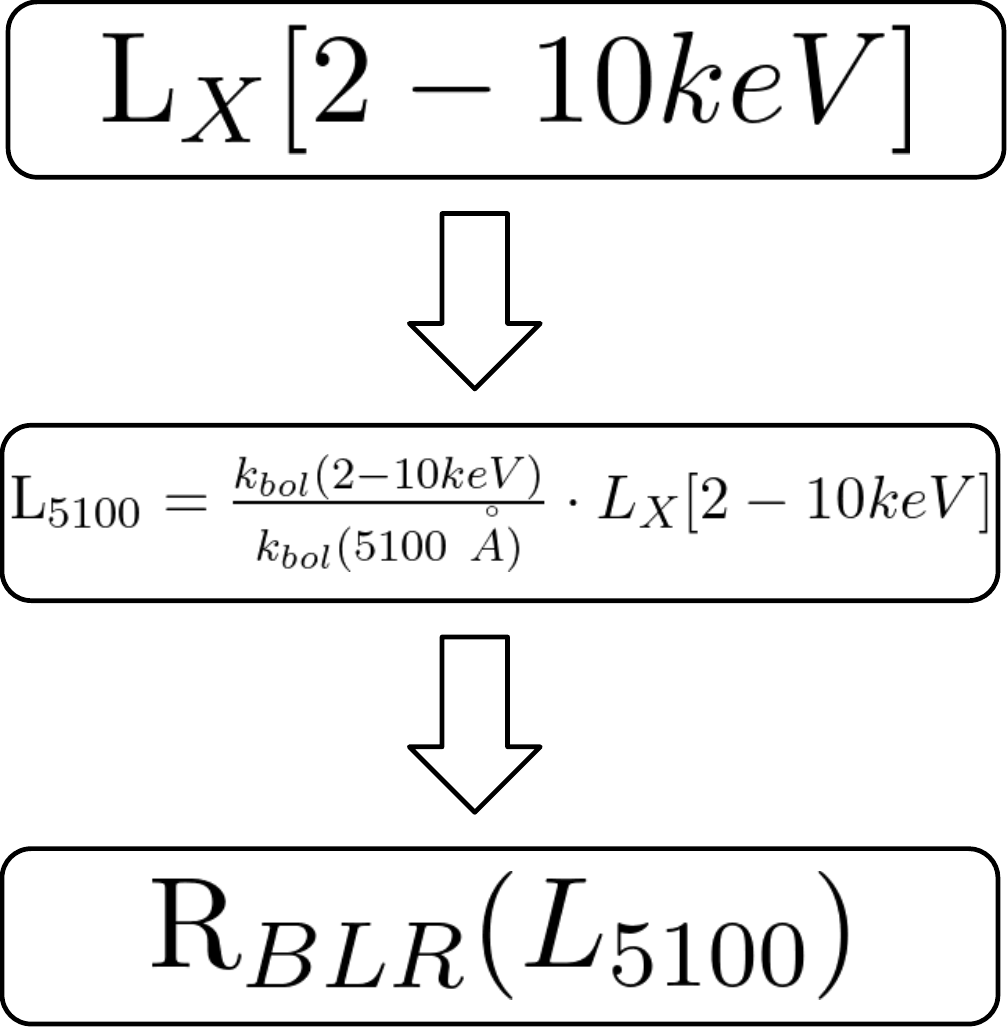}
\caption{General pipeline for deriving H$\beta$~\citep{2013ApJ...767..149B} and $H\alpha$~\citep{2023ApJ...953..142C} BLR radii in this appendix. \label{fig: r_blr_from_xray}}
\end{figure}

\begin{table}
\centering
\caption{Black hole mass estimations using bolometric calibration $L_{5100}-L_{X}$ and $R-L$ relations. \label{tab:MBHs_through_LX_gaussian}}
\begin{tabular}{lccc}
 SSP Model & $\mathrm{FWHM_{H\alpha}}$ & $\mathrm{M_{BH,~R_{B13}}}$ & $\mathrm{M_{BH,~R_{C23}}}$  \\ 
\hline
\hline
  & $\mathrm{km/s}$ & $\times10^{6}\mathrm{M_{\odot}}$ & $\times10^{6}\mathrm{M_{\odot}}$  \\ 
 \hline
\hline
\smallskip
 E-MILES & $ 705.03 \pm 24.49 $ & $ 0.12_{-0.05}^{+0.07} $ & $ 0.11_{-0.05}^{+0.12} $\\
\smallskip
 MILES(+MgFe) & $ 709.74 \pm 24.49 $ & $ 0.13_{-0.05}^{+0.07} $ & $ 0.11_{-0.06}^{+0.12} $\\
\smallskip
 MILES & $ 718.22 \pm 24.96 $ & $ 0.13_{-0.05}^{+0.08} $ & $ 0.11_{-0.06}^{+0.12} $\\
\smallskip
 PEGASE & $ 705.03 \pm 24.73 $ & $ 0.12_{-0.05}^{+0.07}$ & $ 0.11_{-0.05}^{+0.12} $\\
\smallskip
 X-Shooter & $ 690.67 \pm 24.25 $ & $ 0.12_{-0.04}^{+0.07} $ & $ 0.10_{-0.05}^{+0.12} $\\ 
\hline
\end{tabular}
\tablefoot{
$M_{BH,~R_{B13}}$ uses $R_{\mathrm{H\beta}}(L_{5100})$-relation from \citet{2013ApJ...767..149B}, $M_{BH,~R_{C23}}$ uses $R_{\mathrm{H\alpha}}(L_{5100})$-relation from \citet{2023ApJ...953..142C}.
Monte Carlo sampling was used for error propagation, incorporating all errors and inner scatters reported in the relation coefficients. BH mass estimators are calculated as the median, and BH mass error bars are calculated as a 68\% confidence interval. The Gaussian profile is assumed for broad emission lines in the full-spectrum fitting.
}
\end{table}

In this appendix, we discuss the single-epoch BH mass estimation using X-ray luminosity as a proxy of the BLR radius.

To estimate the H$\beta$ BLR radius, we took Eq.~\ref{eq: bolometric_cor} from \citet{2013peag.book.....N} as a relation between $L_{5100}$ and $L_{X,~2-10~keV}$, the relation \ref{eq: hbeta_r_blr_l5100} between $H\beta$-determined $R_{BLR}$ and $L_{5100}$ and obtained $R_{BLR,~H\beta} =  2.16^{+1.34}_{-0.83}~[\mathrm{lt-days}]$. Then we used the virial BH mass definition \ref{eq: virial_mass} with the virial factor $f=1$.
This pipeline builds the $M_{BH,~H\beta}$ (equal to Eq.~\ref{eq: mbh_hbeta}, the only difference is assumption on how to calculate $R_{BLR,~H\beta}$).
We notice that our full-spectrum fitting considers the same FWHM for all broad parts of emission lines; therefore, $\mathrm{FWHM_{H\alpha}=FWHM_{H\beta}}$.

To calculate H$\alpha$ BLR radius defined by \cite{2023ApJ...953..142C}, we used the following equation:
\begin{equation}
\log_{10} \left( \frac{\tau_{\mathrm{H}\alpha}}{1~\mathrm{day}} \right)= (1.59 \pm 0.05) + (0.58 \pm 0.04)\cdot \log_{10} \left( \frac{\lambda L_{\lambda}(5100\mathrm{\AA})}{10^{44}\mathrm{erg\,s}^{-1}} \right),
\label{eq: rhalpha_l5100}
\end{equation}
with intrinsic scatter $\sigma_{int} = 0.31~\mathrm{dex}$,
and obtained $R_{BLR,~H\alpha} = 1.10^{+1.27}_{-0.59}$~[lt-days].
To calculate H$\alpha$ BH mass, we use Eq.~\ref{eq: mbh_ha_virial}.
We adopted the mass of the Sun $M_{\odot} = 1.988416 \times 10^{30}$~[kg] from \cite{2016AJ....152...41P}.

For uncertainty propagation, we used Monte Carlo sampling, incorporating all reported coefficient uncertainties and inner scatters of empirical relations.
We calculate BH masses as the median of the Monte Carlo sampled distribution and uncertainties within a 68\% confidence interval using both H$\beta$ and H$\alpha$ BLR radii and demonstrate them in Table~\ref{tab:MBHs_through_LX_gaussian}. They agreed within 68\%  confidence intervals.

\section{Prediction for BH mass from $M_{BH}(\sigma_{*})$ scaling relation}
\label{sec: appendix_C_mbhs_sigma}
In this appendix, we discuss the $M_{BH}(\sigma_{*})$ relation and the derived upper limit of the BH mass based on two empirical relations.

\citet{2013ARA&A..51..511K} provides $M_{BH}(\sigma_{*})$ scaling relation for classical bulges and elliptical galaxies: 
\begin{equation}
   M_{BH,~KH13}/10^{9}M_{\odot} = \left(0.309^{+0.037}_{-0.033}\right)\cdot\left(\frac{\sigma_{*}}{200~\mathrm{km~s}^{-1}}\right)^{4.38\pm0.29}
\end{equation}
with intrinsic scatter $\sigma_{int}=0.29$.

\citet{2025ApJ...978..115W} reported an updated scaling relation for bulge-aperture $\sigma_{*}$ estimation, using IFU spectral cubes (MUSE, KCWI) fits:
\begin{equation}
   \log_{10}\left(\frac{M_{BH,~W25}}{M_{\odot}}\right) = (7.90\pm0.16) + (2.53\pm0.73)\cdot\log_{10}\left(\frac{\sigma_{*}}{200~\mathrm{km~s}^{-1}}\right)
\end{equation}
with intrinsic scatter $\sigma_{int}=0.47$.

We applied these relations to SDSS-derived $\sigma_{*}$. We discarded the E-MILES \citep{2016MNRAS.463.3409V}, MILES(+MgFe), MILES \citep{2010MNRAS.404.1639V}, and PEGASE \citep{2004A&A...425..881L} SSP full-spectrum fitting results for the SDSS spectra, since the minimizer converged on $\sigma_{*}$ < LSF/3. 

For the uncertainty propagation, we used Monte Carlo sampling, incorporating all reported coefficient uncertainties and inner scatters of empirical relations.
We calculate BH masses as the median of the Monte Carlo sampled distribution and uncertainties within a 68\% confidence interval. The predicted BH masses by both relations are demonstrated in Table~\ref{tab: mbh_sigma_stellar} and agree within 68\% confidence intervals.

\begin{table}[b]
\centering
\caption{BH mass predictions using $M_{BH}(\sigma_{*})$ relations. \label{tab: mbh_sigma_stellar}}
\begin{tabular}{lccc}
\smallskip
SSP Model & $\sigma_{*}$ & $\mathrm{M_{BH,~KH13}}$ & $\mathrm{M_{BH,~W25}}$  \\
\hline
\hline
& $\mathrm{km/s}$ & $\times10^{6}\mathrm{M_{\odot}}$ & $\times10^{6}\mathrm{M_{\odot}}$\\
\hline
\hline
\smallskip
X-Shooter & $ 67.3 \pm 8.0 $ & $ 2.64^{+2.59}_{-1.34} $ & $ 5.35^{+13.21}_{-3.78}$\\
\hline
\hline
\end{tabular}
\tablefoot{
$\mathrm{M_{BH,~KH13}}$ corresponds to $\mathrm{M_{BH}(\sigma_{*})}$-relation taken from \citet{2013ARA&A..51..511K}, $M_{BH,~W25}$ corresponds to $\mathrm{M_{BH}}(\sigma_{*})$ relation taken from \citet{2025ApJ...978..115W}. Monte Carlo sampling was used for error propagation, incorporating all errors and inner scatters reported in the relation coefficients. BH mass estimators are calculated as the median, and BH mass error bars are calculated as a 68\% confidence interval. The Gaussian profile is assumed for broad emission lines in the full-spectrum fitting.
}
\end{table}

\end{appendix}
\end{document}